\documentclass[iop]{emulateapj}

\usepackage{graphicx}
\usepackage{adjustbox}

\slugcomment{Submitted to ApJ}

\shorttitle{Spectroastrometric Detection of Exomoons}
\shortauthors{Agol, Jansen, Lacy, Robinson \& Meadows}

\begin{document}

\title{The Center of Light: Spectroastrometric Detection of Exomoons}

\author{Eric Agol\altaffilmark{1,2}, Tiffany  Jansen, Brianna Lacy, Tyler D. Robinson\altaffilmark{2,3} \and Victoria Meadows\altaffilmark{1,2}}
\email{agol@uw.edu}

\affil{Astronomy Department, University of Washington, Seattle, WA 98195}

\altaffiltext{1}{University of Washington Astrobiology Program}
\altaffiltext{2}{NASA Astrobiology Institute's Virtual Planetary Laboratory, Seattle, WA 98195, USA}
\altaffiltext{3}{NASA Ames Research Center, MS 245-3, Moffett Field, CA 94035, USA}

\begin{abstract}
Direct imaging of extrasolar planets with future space-based
coronagraphic telescopes may provide a means of detecting companion moons at
wavelengths where the moon outshines the planet.  We propose
a detection strategy based on the positional variation of the center of
light with wavelength, ``spectroastrometry." This new application
of this technique 
could be used to detect an exomoon, to determine the exomoon's 
orbit and the mass of the host exoplanet, and to 
disentangle of the spectra of the planet and moon. We consider
two model systems, for which we discuss the requirements for detection
of exomoons around nearby stars. We simulate the
characterization of an Earth-Moon analog system with spectroastrometry,
showing that the orbit, the planet mass, and the spectra of both bodies can be
recovered. To enable the detection and characterization of exomoons
we recommend that coronagraphic telescopes should
extend in wavelength coverage to 3 micron, and should be designed with 
spectroastrometric requirements in mind.
\end{abstract}


\section{Introduction}

Techniques and ideas for detecting moons orbiting exoplanets 
have progressed rapidly in the last decade.  These advances 
are driven, at least in part, by a desire to extend our 
understanding of satellite formation and origins to cases 
beyond the Solar System.  For example, while the formation 
of moons and/or moon systems may be common around giant 
planets, there exist many physical models for this process 
\citep{1982Icar...52...14L,2002AJ....124.3404C,
2003Icar..163..198M,2012ApJ...753...60O}.  Additionally, 
the likelihoods of satellite formation by giant impact 
\citep{1975Icar...24..504H,1976LPI.....7..120C} and 
capture \citep{1966AJ.....71..585M,1984Natur.311..355M} 
remain uncertain \citep[e.g.,][]{2004ApJ...613L.157A,
2006Natur.441..192A}.

Exomoons can be potentially habitable targets in their own right, 
and may also affect the habitability and characterization of 
the host planet \citep{2014AsBio..14..798H}.  Habitable, 
Earth-like moons orbiting gas 
giant planets in circumstellar habitable zones have 
been proposed \citep{1997Natur.385..234W,
2000ESASP.462..199K} and discussed 
\citep{2006ApJ...648.1196S,2012A&A...545L...8H,
2013AsBio..13...18H,2015A&A...578A..19H,2015IJAsB..14..335H,2013ApJ...774...27H,2013MNRAS.432.2994F,
2010ApJ...712L.125K,HellerArmstrong2014,Reynolds1987}.  For rocky planets in the habitable 
zone, the undetected presence of a large satellite can 
confuse characterization attempts, as the moon is an 
additional source of thermal or reflected light 
\citep{2011ApJ...741...51R}, potentially with its own 
molecular features \citep{2014PNAS..111.6871R}.  
Detecting large companions could also aid in the 
characterization of potentially habitable exoplanets, as 
large satellites can help provide long-term obliquity, and 
thus climate, stability \citep{1993Natur.361..615L}.
However, the presence of large moons does not necessarily 
imply such stability \citep{2002LPI....33.2017W}, nor
 is a moon necessitated for moderate obliquity stability
\citep{Lissauer2012}, and it might reduce habitability
near the outer edge of the insolation habitable zone
\citep{Armstrongetal2014}.

Given the importance of satellites to exoplanet 
characterization and in advancing our understanding of 
planet formation models, it is not surprising that many  
techniques for exomoon detection have been proposed.  Most 
techniques have focused on analyzing exoplanet transit timing 
and/or duration signals \citep{1999A&AS..134..553S,
2006A&A...450..395S,2007A&A...470..727S,2009MNRAS.392..181K}, 
which can vary due to the orbit of the planet about the 
planet-moon barycenter.  Additionally, mutual events, with a 
moon occulting its host star or planet while the planet 
transits the star, as well as the overlap of the moon
and planet during transit, can  cause photometric variations in the 
stellar signal \citep{1999A&AS..134..553S,2006A&A...450..395S,2011MNRAS.416..689K,
2011ApJ...743...97T,Heller2014}, while
spectroscopy during the transit of an exomoon that is well
separated from its companion planet might allow for
a measurement of the exomoon's atmospheric transmission
spectrum \citep{2010ApJ...712L.125K}.
Most of these effects have been 
considered in an ongoing search for exomoons in the 
\textit{Kepler} dataset \citep{2012ApJ...750..115K}.

By comparison, relatively few techniques have been developed  
that are relevant to direct observations of exoplanets, where 
the planet is resolved from the host star in either reflected 
or emitted light.  Development of such techniques is prudent 
given the near- and long-term interest in exoplanet direct 
imaging and characterization \citep{2014arXiv1401.3741K,2014SPIE.9143E..2KS,2015arXiv50303757S}.  
\citet{2007A&A...464.1133C} discussed using mutual transit 
and shadowing effects to detect a moon in an unresolved 
planet-moon system, which is made difficult by requiring an 
observational duty cycle near to 100\% of the moon's orbit
and/or a fortuitous inclination of the moon's orbit relative to the observer.  
\citet{2009AsBio...9..269M} discussed the influence of a 
large moon on the bolometric thermal-infrared phase curve 
of its host planet, and \citet{2011ApJ...741...51R} suggested 
that phase-dependent variability at wavelengths corresponding 
to strong absorption bands for the host planet could indicate 
the presence of an airless moon.  Recently, 
\citet{2013ApJ...769...98P} discussed the direct detection of 
tidally-heated exomoons, which can be made to be very luminous 
depending on the amount of input tidal energy. 

A major obstacle to exomoon detection in direct 
observations of exoplanets is angular resolution.  For example, at a distance of 1.34~pc ($\alpha$ Cen system) 
the Earth-Moon angular separation would span 1.9~mas, and 
the Saturn-Titan angular separation at that same distance would 
be 6.0~mas.  Thus, to resolve these pairs of worlds, a telescope 
operating in V-band would need to have a diameter of about 
25~m or more, and this value increases to be on the order of a kilometer at 10~$\mu$m in the thermal infrared.

Here, we turn to a mature technique in observational 
astronomy, spectroastrometry \citep{Bailey1998,2008LNP...742..123W}, as 
a new means for detecting exomoons in direct observations 
of exoplanets.  In spectroastrometry, the wavelength-dependent 
shift in the light centroid of an unresolved source can be 
used to indicate the presence of an unresolved companion body and to 
yield orbital parameters of the companion.  This is limited
by the optical design and diffraction within the telescope, 
which spread a point source of light into a broadened distribution
of photons, referred to as the ``point spread function," or PSF.
A spectroastrometric detection is made possible by the angular 
shift of the PSF {\it centroid} with wavelength, which can be measured to 
precisions better than the angular resolution of the telescope, 
which sets the size of the PSF, yielding information at the angular 
scale of the exoplanet-exomoon separation.

\begin{figure}
\centering
\includegraphics[width=\hsize]{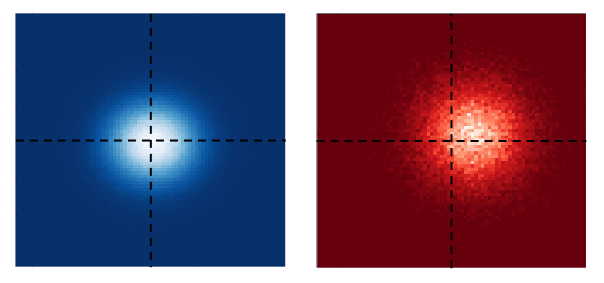}
\caption{Spectroastrometric signal. Left: at a bluer wavelength
dominated by the planet, the centroid aligns with the planet's position
at the origin.  Right: at a redder wavelength dominated by the 
moon (within an absorption band in which the planet's spectrum is
dark), the centroid shifts to the position of the moon.}
\label{fig:spectroastrometric_signal}
\end{figure}

We propose a new technique to detect exomoons using a high-contrast
imaging coronagraph which suppresses the light of the host
star exterior to an inner working angle \citep[e.g.][]{Guyon2006,2014SPIE.9143E..2KS}, 
enabling direct imaging 
of the unresolved exoplanet and exomoon. This can alternately
be accomplished with an external star shade \citep{cash2006}.
At wavelengths where the exoplanet
dominates the flux, the centroid will approach the position of the
exoplanet, while at wavelengths where the exomoon dominates, the
centroid will approach the position of the exomoon (Figure \ref{fig:spectroastrometric_signal}).
Spectroastrometry then allows for the
measurement of the centroid shift of the PSF containing the exoplanet and
its moon, resulting in the detection of the exomoon, as well
as enabling follow-up observations for characterization of
the star-planet-moon system.  This technique works optimally for
an instrument that is capable
of measuring both spatial and spectral information simultaneously,
although we do not specify the detector design in this study.

We begin this paper with a discussion of our spectral models for
the planets and moons (\S \ref{spectral_models}).
Next, we discuss the methods of our simulations (\S \ref{methods}), starting with
the fiducial model parameters of the systems we study (\S \ref{ms parameters}),
the definition of spectroastrometric signal (\S \ref{spectroastrometry}), 
the computation of the signal-to-noise
for detection of a moon (\S \ref{SNR}), and finishing with a description of
Monte Carlo simulations of spectroastrometric observations (\S \ref{monte_carlo}).  Next, we present results of the
optimization of wavelength choice for detection (\S \ref{ms}), and an example
of the characterization of a hypothetical nearby planet-moon system with
spectroastrometry (\S \ref{earth_moon_characterization}).
We  discuss how our results will scale with the various
assumed parameters (\S \ref{telescope parameters}).  We end with a discussion and conclusions.

\section{Description of Spectral Models}
\label{spectral_models}

We use two fiducial model systems to explore the prospects for spectroastrometry:
an Earth-Moon twin around the star Alpha Centauri A 
($\alpha$ Cen A; 1.34 parsecs), and an Earth-Jovian 
analog at 1 AU around a G2V (sun-like) star at 10 parsecs (Table \ref{table01}).  
We use three distinct 
planet spectral models to simulate the spectra of an exo-Earth, an exo-Moon, 
and a warm Jupiter (\emph{i.e.,} a Jupiter-like planet at an orbital distance of 
1-2 AU from a Sun-like star).  Our Earth and Moon models are described in the 
following subsections.  The spectrum of the warm Jupiter was generated by the 
radiative-convective models described in \citet{burrowsetal04} and 
\citet{sudarskyetal03}.  These models produce phase-averaged spectra of 
irradiated gas giants for a set of assumed elemental abundances, and 
include the effects of cloud condensation 
and multiple scattering.  The atmosphere in these models is assumed to be planar, and 
the particular model that we use in this work is for a solar-composition, 
Jupiter-mass planet orbiting a G2V star at a distance of 1 AU.  The jovian's 
atmosphere is taken to be in chemical equilibrium, and the equilibrium 
abundances of key atmospheric trace gases are computed according to 
\citet{burrows&sharp99}.

\begin{table*}
\begin{center}
\caption{The moon and planetary parameters used in our model, where $D_{tele}$ is the telescope diameter, \textit{d} is the distance from the observer, $t_{obs}$ is the duration of exposure, $\epsilon$ is the telescope efficiency factor.\label{table01}}
\begin{adjustbox}{width=1\textwidth}
\small

    \begin{tabular}{l|ccccccccc}
            & Moon radius &  Planet radius   & Planet-moon & Orbital & $D_{tele}$ & \textit{d} & $t_{obs}$ & $\epsilon$ & Semi-major     \\ 
          System & (m) & (m) & separation (m) & period (d) & (m) & (pc) & (hr) & & axis (AU) \\ \cline{1-10} \\
    Earth-Moon   &  1.738$\times 10^{6}$ & 6.371$\times 10^{6}$ &  3.844$\times 10^{8}$ & 27.32 & 12 & 1.34 & 24 & 0.2 & 1.23   \\[.1cm]
    Jovian-Earth &  6.371$\times 10^{6}$ & 6.991$\times 10^{7}$ & 3.064$\times 10^{9}$ & 34.60 & 12 & 10 & 24 & 0.2 & 1     
    \end{tabular}
\end{adjustbox}
\end{center}
\end{table*}

\subsection{Earth Spectrum Model}

To simulate Earth's disk-integrated spectrum, we use the NASA Astrobiology 
Institute's Virtual Planetary Laboratory three-dimensional, multiple scattering spectral Earth model, which 
generates temporally and spectrally resolved disk-integrated synthetic observations   
of Earth. This model has been described and extensively validated both for temporal 
variability and for a variety of phases, at wavelengths from the near-ultraviolet 
through the IR in previous papers \citep{robinsonetal10,robinsonetal11,robinsonetal14}, 
so only a brief description of the model is presented here.  

In our simulations, we divide Earth into a number of equal area pixels 
according to the HEALPix scheme \citep{gorskietal05}, and Earth's 
disk-integrated spectrum is computed by summing the radiances coming from 
the pixels on the observable hemisphere. The wavelength-dependent radiance 
coming from any given pixel is assembled from a lookup table that contains 
spectra generated over a grid of different solar and observer zenith and 
azimuth angles. Elements within the lookup table, which are generated using a 
one-dimensional, line-by-line radiative transfer model \citep{meadows&crisp96}, 
are computed for a variety of different surface and atmospheric conditions, as 
well as several different cloud coverage scenarios (\emph{e.g.}, thick, low 
cloud or thin, high cloud).

To simulate time-dependent changes in Earth's spectrum we use spatially-resolved, 
date-specific observations of key surface and atmospheric 
properties from Earth observing satellites as input to our Earth model. 
Gas mixing ratio and/or temperature profiles are taken from the Microwave Limb 
Sounder \citep{watersetal06}, the Tropospheric Emission Spectrometer 
\citep{beeretal01}, the Atmospheric Infrared Sounder \citep{aumann03}, and the 
CarbonTracker project \citep{petersetal07}.  Snow cover and sea ice data as 
well as cloud cover and optical thickness data are taken from the Moderate 
Resolution Imaging Spectroradiometer instruments \citep{salomonsonetal89} aboard 
NASA's Terra and Aqua satellites \citep{halletal95,riggsetal95}. 
Wavelength-dependent optical properties for liquid water clouds were derived using a Mie 
theory model \citep{crisp97} and were parametrized using geometric optics for 
ice clouds \citep{muinonenetal89}.  The spectra presented in this work are for 
Earth at northern vernal equinox at quadrature phase.

\subsection{Moon Spectrum Model}

We divide the Moon's spectrum into two components: reflected solar and 
emitted thermal.  At quadrature, the reflected component dominates at 
wavelengths below 3.5~$\mu$m and the emitted component dominates at 
wavelengths above this.  The thermal component of the spectrum is 
computed using the model presented in \citet{2011ApJ...741...51R}.  The  
model generates phase-dependent, thermal infrared spectra of the Moon 
assuming globally-averaged values of the lunar Bond albedo, nightside 
temperature, and the spectrally-resolved surface emissivity.

The lunar phase function is markedly non-Lambertian, so we use empirical 
models of the Moon's phase-dependent reflectivity to simulate the 
shortwave component of the lunar spectrum.  Spectrally-resolved measurements 
of the lunar surface phase function from the RObotic Lunar Observatory (ROLO) 
are used to simulate the reflected light component of the lunar spectrum at 
star-planet-observer angles (\emph{i.e.}, phase angles) between 0$^{\circ}$ 
and 97$^{\circ}$, which are the phase angles for which the surface phase 
function has been published \citep{burattietal11}.  The ratio of the 
specific intensity emerging from a surface element 
($I_{\lambda}(\alpha,\mu_{0},\mu)$, where $\alpha$ is the phase angle, 
and $\mu_{0}$ and $\mu$ are the cosines of the incident solar angle and 
the emission angle, respectively) to the incident specific solar 
flux ($F_{\lambda} = F_{\sun}/\pi$) is given by \citep{chandrasekhar60}, 
\begin{equation}
I_{\lambda}(\alpha,\mu_{0},\mu)/F_{\lambda} = f(\alpha)[\mu_{0}/(\mu+\mu_{0})] \ ,
\label{eqn:moon1}
\end{equation}
where $f(\alpha)$ is the surface solar phase function (which is distinct from 
the planetary phase function), and the final collection of terms in brackets 
provides the functional form of the lunar scattering law.  The ROLO data 
provide $f(\alpha)$, so that we can integrate $I_{\lambda}(\alpha,\mu_{0},\mu)$ 
from Eq.~\ref{eqn:moon1} over the illuminated disk, which eliminates the 
dependence on $\mu_{0}$ and $\mu$, and allows us to compute the phase-dependent, 
reflected light spectrum of the Moon, $I_{\lambda}(\alpha)$.

At phase angles larger than 97$^{\circ}$, where $f(\alpha)$ is not reported, we 
model the Moon's reflected light spectrum using the lunar phase functions of 
\citet{lane&irvine73}, who measured the Moon's brightness over a wide range 
of phases through a series of broadband filters from the near-ultraviolet to 
the near-infrared.  By pairing these measurements of the Moon's phase function 
with a medium-resolution ($\lambda/\Delta\lambda \sim$~500) measurement of the 
disk-integrated lunar spectrum from NASA's EPOXI mission \citep{Livengood2011}, 
we can infer the Moon's spectrum at phase angles other than that at which the 
EPOXI data were acquired.  Thus, for large phase angles, we take the 
phase-dependent, disk-integrated specific brightness of the Moon to be
\begin{equation}
I_{\lambda}(\alpha) = I_{\lambda}(\alpha^{\prime})\frac{\Phi(\alpha)}{\Phi(\alpha^{\prime})} \ ,
\label{eqn:moon2}
\end{equation}
where $\alpha^{\prime}$ is the phase angle of the EPOXI observations, and $\Phi$ 
is the planetary phase function measured by \citet{lane&irvine73}. The lunar EPOXI 
observations span 0.3~$\mu$m to about 4.5~$\mu$m in wavelength, and were taken at 
a phase angle of 75.1$^{\circ}$.  In general, our two approaches to simulating 
the reflected-light component of the Moon's spectrum agree to within measurement 
error at phase angles near quadrature, $\alpha = 90^\circ$, which is the phase that we emphasize in 
this work.

\section{Methods}
\label{methods}

The spatial resolution needed to detect the spectroastrometric signal for our
fiducial simulated planet-moon systems is much greater than any telescope can 
currently provide, and greater than the proposed first-generation direct
imaging missions.  Both planet-moon systems have a separation of $\approx 2$
milliarcseconds (mas), while the Hubble Space Telescope has a PSF width of
$\approx$100 mas at 2.5 micron (and no coronagraph), and first-generation imaging 
telescopes will have similar spatial resolutions.  Consequently we focus on 
future second-generation telescopes.  To simulate a spectroastrometric 
detection of the exomoon models presented in this paper, we 
explored the multidimensional parameter space of an idealized
coronagraphic space telescope capable of such a feat. 
A space telescope would be ideal for exomoon detection due to the ability to 
observe in the Earth's atmospheric windows and to avoid atmospheric turbulence 
without the aid of adaptive optics;  these same requirements apply to coronagraphic 
detection and characterization of Earth-like planets in their habitable zones. 
In \S{\ref{ms parameters}} we outline the parameters used 
in our model, in \S{\ref{spectroastrometry}} we define the spectroastrometric
signal, in \S{\ref{SNR}} we discuss the photon-limited noise and
signal-to-noise, and in \S{\ref{monte_carlo}} we describe Monte Carlo simulations of spectroastrometric 
observations.

\subsection{Model System Parameters}
\label{ms parameters}

We have assumed a planet-moon system whose moon orbits in the planet's equatorial plane in prograde
motion. The planet-moon separation for the ExoEarth-Moon system was taken to be the current Earth-Moon separation, 
while the separation for the Jovian-Earth system was taken to be 30$\%$ of the Hill radius of a warm 
Jupiter at 1 AU from a Sun-like star (corresponding to 43 $R_{Jupiter}$), slightly less than the critical semi-major axis for a prograde satellite
\citep{barneso'brien02}. For the preliminary exploration of the signal-to-noise
ratio for different spectral resolutions and wavelength pairs, we set the
telescope diameter to 12 meters, distance to the system from the observer to 1.34
parsecs (Earth-Moon) and 10 pc (Jovian-Earth), and exposure time to 1 day (Table \ref{table01}).
We assume that the moon-planet position is fixed over the range of the exposure;
since both of our fiducial systems have orbital periods of $\approx 30$ days,
the centroid will shift by $\approx 12^\circ$ over this time, which is only
20\% of the planet-moon separation, and a very small fraction of the width
of the PSF.
We have selected a 12-meter telescope based on the range of apertures
being considered for the ``High-Definition Space Telescope"  \citep[HDST;][]{Dalcanton2015}, which is informed by a (conservative) estimate of the detectability of Earth-like planets \citep{Stark2015}.
The telescope efficiency factor, which accounts for instrument throughput, detector 
quantum efficiency, and photometric aperture losses, was kept at 20$\%$ throughout 
all calculations. These system parameters may be found in Table \ref{table01}.
Section \S{\ref{telescope parameters}} explores the effect that varying the 
telescope size and distance from the observer has on the signal to noise ratio.

Such a large telescope would be capable of rapid detection of Earth-sized
exoplanets (in of order a few weeks) with a direct-imaging survey of 
nearby Sun-like stars \citep{agol2007}.
The planets found with this initial coronagraphic imaging survey could then be
prioritized based upon, for example, their presence within their host-star
habitable zone and potential for the existence of a large, detectable exomoon, 
to motivate the additional observing time for detailed
spectral and spectroastrometric characterization.

We note that the parameters of the Jovian-Earth model system we have chosen
are driven by favorable observational detectability rather than theoretical
prejudice. Models of satellite formation around giant planets have been tailored
for the Solar System to reproduce, for example, the mass ratio of $10^{-4}$
observed for the satellite systems of the giant planets \citep{2002AJ....124.3404C,
Sasaki2010}, so that giant planets may harbor satellites as large as $0.7R_\oplus$
\citep{2015ApJ...806..181H}.  In addition, models of in-situ formation of satellites from
a disk produce regular satellites at distances of $6-30 R_{Jupiter}$.
Thus, within the context of these models, an Earth-sized satellite at 43 $R_{Jupiter}$ 
distance from a giant planet is unexpected, but may have formed
by other means, including capture and migration.

\subsection{Spectroastrometric Signal}
\label{spectroastrometry}

The spectroastrometric signal is the difference in the centroid positions when the 
system is observed at two different wavelengths.  We define the centroid position,
${\bf c}(\lambda)$, as being the flux-weighted center of light of the planet and moon as 
a function of wavelength, $\lambda$, in the direction of two angular sky coordinates, 
$(x,y)$, which we denote as ${\bf c}(\lambda) = (c_x(\lambda),c_y(\lambda))$,
where $c_{x,y}$ are the $(x,y)$ centroids with respect to the origin of the star-planet-moon
system (and the orientation of the coordinates is chosen with respect to an
observer-defined reference direction).  Note that the angular coordinates $(x,y)$ are expressed in milliarcseconds
(mas), which is the typical order of the moon-planet separation.  

The detector will be capable of measuring this signal over a wave band with
central wavelength $\lambda$ and spectral resolution $R$, giving the width
of the band $\Delta \lambda = \lambda/R$.  We define the band-averaged centroid
as
\begin{equation}
\bar{\bf c}(\lambda) = \frac{\int_{\lambda -\Delta \lambda/2}^{\lambda +\Delta \lambda/2} {\bf c}(\lambda) \lambda F_\lambda d\lambda}{\int_{\lambda -\Delta \lambda/2}^{\lambda +\Delta\lambda/2}  \lambda F_\lambda d\lambda},
\end{equation}
where $F_\lambda$ is the total flux of the planet and moon, and we have weighted
the centroid by the number of photons detected.

In computing the model centroid, we define $F_{m,\lambda}$ and $F_{p,\lambda}$ to represent the flux of the moon and the planet, respectively, and ${\bf r}_m$ and ${\bf r}_p$ to represent the positions of these bodies (in two dimensions, projected onto the sky plane, perpendicular to our line of sight). The position of the wavelength-dependent centroid, ${\bf c}(\lambda)$, is
\begin{eqnarray}
{\bf c}({\lambda}) &=& \frac{F_{m,\lambda}{\bf r}_m + F_{p,\lambda} {\bf r}_p}
{d(F_{m,\lambda} +  F_{p,\lambda})}\cr
&=&\frac{F_{m,\lambda}\vec{\beta}}{F_{m,\lambda} +  F_{p,\lambda}} + \frac{{\bf r}_p}{d}
\end{eqnarray}
where $\vec{\beta}=({\bf r}_m-{\bf r}_p)/d$ is the angular sky separation between
the moon and planet in radians at a distance $d$ from the observer (we convert
this to milliarcseoconds).  Note that this equation neglects the variation
of the center of light of each body with the illumination phase;  this is on the 
order of the angular size of each body, which is much smaller than the
spectroastrometric signal.

Measuring ${\bf c}({\lambda})$ requires a reference position on the sky, for example
the planet's position, while the planet's position is not known a priori in
the presence of a centroid-shifting moon. However, knowledge of the planet's
position is not needed when considering only the difference between the centroids
measured in different spectral bands. This is the spectroastrometric signal, $S(\lambda_1,\lambda_2)$,
which scales as the scalar difference of the planet/moon centroid in two
wave bands,
\begin{equation}
S(\lambda_1,\lambda_2)= \vert\bar{\bf c}(\lambda_{1}) - \bar{\bf c}(\lambda_{2})\vert,
\end{equation}
which we measure in milliarcseconds.
Note that it may be possible to select a different resolution $R$ for
each wave band.
A significant advantage
of measuring the spectroastrometric signal, $S(\lambda_1,\lambda_2)$, rather than ${\bf c}(\lambda)$ is that
the {\it absolute} position of the planet-moon system does not need to be calibrated;
only the relative position with wavelength needs to be measured.  This obviates
the need for an absolute sky coordinate reference frame, which can be a challenging
measurement to make (although the host star could perhaps be used, as discussed
below).  Also, the position of the planet is affected by its orbit about
the star, its illumination, its orbit about the center of mass with the moon, 
and its gravitational perturbation by companion planets;  measurement of the
spectroastrometric signal eliminates all of these confounding effects.
Note that this technique does not require flux calibration since the spectroastrometric
signal is normalized by the total flux of the planet and moon.

\subsection{Signal to Noise Ratio of Detection}
\label{SNR}

The spectroastrometric noise scales as the ratio of the width of the point spread
function incident on the detector to the square root of the number of photons,
assuming Poisson noise from the planet and moon dominates the uncertainty. In light
of the extraordinary engineering called for in the success of this project, we have
assumed that an ideal coronagraphic space telescope would produce negligible
instrumental noise (due to the dark current, read noise, scattered light, pixel
size, etc). We estimate the number of photons, $N(\lambda)$,
incident on the coronagraph's detector due to the planet and moon 
within a band of width $\Delta \lambda$
for some exposure time $t_{obs}$ to be
\begin{eqnarray}
N(\lambda) &=& \pi\epsilon\frac{D^{2}_{tele}}{4} t_{obs} \int_{\lambda-\Delta
\lambda/2}^{\lambda + \Delta \lambda/2} \frac{\lambda}{hc} F_\lambda d\lambda \cr
&\approx& \pi\epsilon\frac{D^{2}_{tele}}{4}\frac{\Delta\lambda}{\lambda}
\frac{\lambda^{2}}{hc}{t_{obs}}F_{\lambda},
\end{eqnarray}
where $\epsilon$ is the chosen efficiency factor of the telescope (defined
as the fraction of all photons incident on the telescope aperture that are
measured by the detector; we assume that it is wavelength-independent), $D_{tele}$ 
is the diameter of the telescope, $h$ the Planck constant, $c$ the speed of light,
and $F_{\lambda}$ is the wavelength dependent flux density (the approximation assumes that
this flux is roughly constant across a narrow band). A deeper analysis of this
calculation can be found in \citet{agol2007};
note that $R=\lambda/\Delta \lambda$ is the spectral resolving power.  

We assume
that the PSF of the planet is well approximated by an Airy disk, which we in turn
approximate as a Gaussian with angular profile $I(\theta)\propto \exp{-\frac{1}{2}(\theta/\sigma_{PSF})^2}$,
where $\sigma_{PSF} = 0.45 \lambda/D_{tele}$ is the angular size of the Airy disk,
and $\theta$ is the angular coordinate from the center of the PSF.
The precision of the measurement of the centroid improves with the
square root of the number of photons detected. 
Thus, the spectroastrometric noise for an ideal (photon-noise limited) 
coronagraphic space telescope is then
\begin{equation}
\sigma(\lambda) = \frac{\sigma_{PSF}}{N(\lambda)^{1/2}} 
\approx \frac{0.45\lambda}{D_{tele}}
\left(\pi\epsilon\frac{D^{2}_{tele}}{4}\frac{\Delta\lambda}{\lambda}
\frac{\lambda^{2}}{hc}t_{obs}F_{\lambda}\right)^{-1/2}.
\end{equation}

The spectroastrometric signal-to-noise ratio (hereafter SNR) we define
as 
\begin{equation}
\text{SNR} \equiv \frac{S(\lambda_1,\lambda_2)}{\sqrt{\sigma(\lambda_1)^2
+\sigma(\lambda_2)^2}} 
\end{equation}
between two wavelengths $\lambda_1$ and $\lambda_2$.
Below we set the threshold for exomoon detection in our model systems 
at a minimum 5$-\sigma$ confidence level, i.e.\ $SNR \ge 5$.

In practice, the noise is typically dominated by the moon wave band.
This gives an approximate expression for the signal-to-noise ratio of
\begin{equation}
\text{SNR} \approx \beta \frac{D_{tele}}{0.45\lambda_1}\left(\pi\epsilon\frac{D^{2}_{tele}}{4}\frac{\Delta\lambda_1}{\lambda_1}\frac{\lambda_1^{2}}{hc}{t_{obs}}F_{\lambda_1}\right)^{1/2},
\end{equation}
where we have assumed that $\lambda_1$ is dominated by the moon
while $\lambda_2$ is dominated by the planet, and $\beta=\vert
\vec{\beta}\vert$ is
the moon-planet separation on the sky at the time of observation.  Note
that in the wave band dominated by the moon, $F_\lambda \propto
(R_m/d)^2$, so overall the signal-to-noise scales as
\begin{equation}
\label{snr_scaling_equation}
SNR \propto a_{mp} R_m \left(\frac{D_{tele}}{d}\right)^2 \left(\frac{\epsilon t_{obs}}{R}\right)^{1/2}
\left(\frac{F_{\lambda_1}d^2}{\pi R_m^2}\right)^{1/2},
\end{equation}
where $a_{mp} = \vert {\bf r}_m-{\bf r}_p\vert = \beta d$ 
is the sky-plane projected physical separation of the planet and
moon, $R_m$ is the radius of the moon, and the last quantity
in this equation is the disk-integrated specific intensity of the 
moon.

We use these equations below for estimation of the detectability
of a moon-planet system, while for characterization of a detected system,
we employ Monte Carlo simulations, described next.

\subsection{Monte Carlo Simulations}
\label{monte_carlo}

We have carried out Monte Carlo simulations of coronagraphic observations
of a detected moon-planet system with the same
idealized assumptions: we assume a broad range of wavelength sensitivity,
and we assume that the photon counting noise and diffraction
limit can be achieved.
We define spectral bins equally spaced in log wavelength between
a minimum and maximum wavelength.  For each spectral bin, we compute
the predicted number of photons within the bin for each body, 
$N(\lambda)$, and then 
draw the observed number of photons from a Poisson distribution. 
We next add a random Gaussian deviate in both directions on the sky 
with a standard deviation of $\sigma(\lambda) =0.45\lambda/(D_{tele}N(\lambda)^{1/2})$
at the position of each body.
We then compute the flux-weighted average positions of both bodies
to obtain $\bar{\bf c}(\lambda)$,
and compute the centroid uncertainty, $\sigma(\lambda)$, 
from a weighted mean of the individual bodies' centroid
uncertainties.  The result of this is a total photon flux and $(x,y)$
centroid position for each spectral bin, ${\bf c}(\lambda)$, as well 
as the photon-noise
limited uncertainty (which we assume to be the same in the $x-$ and
$y-$ directions).

We have also run simulations in which we randomly draw individual
photons from the spectral shapes of each body, and assign a position
and positional uncertainty to each photon. We then bin these 
photons in wavelength to measure the spectrum, and
compute the centroid within each bin to measure the spectroastrometric
signal, $\bar{\bf c}(\lambda)$;  we compute uncertainties on the 
photon flux from the square
root of the number of photons in each bin, and on the centroid from
the scatter in the photon positions divided by the square root of
the number of photons in each bin. This approach
is more time intensive, yet gives equivalent results to the pre-binned
spectrum approach.

\section{Results}
\label{results}

Our results outline the range of instrumental, spectroscopic, and planetary system parameters for which we 
expect the spectroastrometric method to detect the presence of an exomoon. In \S{\ref{ms}}, we discuss the 
optimum observing wavelengths and spectral resolutions found with telescope and planet-moon system parameters 
fixed at values described in \S{\ref{ms parameters}}. In \S{\ref{earth_moon_characterization}} we give an example of characterization of an Earth-Moon twin at the distance
of $\alpha$ Cen.
In \S{\ref{telescope parameters}}, we discuss the results 
of varying the distance and telescope size for both systems while spectral resolution and observing wavelengths 
are fixed at favorable values.\\

\subsection{Optimum Spectral Resolutions and Observing Wavelengths for the Model Systems}
\label{ms}

Sections \S{\ref{moon earth results}} and \S{\ref{earth jovian results}} discuss our 
particular model systems in more detail. Results for both systems reflect the inherent 
challenge of spectroastrometry: moon-dominated 
wavelengths tend to exist because the host planet is dim due to atmospheric absorption, not due to
emission features of the moon. Thus, the bands with highest centroid offset inherently have less total flux and an 
elevated level of noise. In order to bring the SNR at these wavelengths above the detection 
threshold, a balance must be struck between a spectral resolution which is low enough to let in
many photons, yet high enough to reveal a measurable shift in the centroid.  

\subsubsection{Moon-like Exomoon Orbiting an Earth-like Planet}
\label{moon earth results}

\begin{figure*}
\centering
\includegraphics[width=14cm]{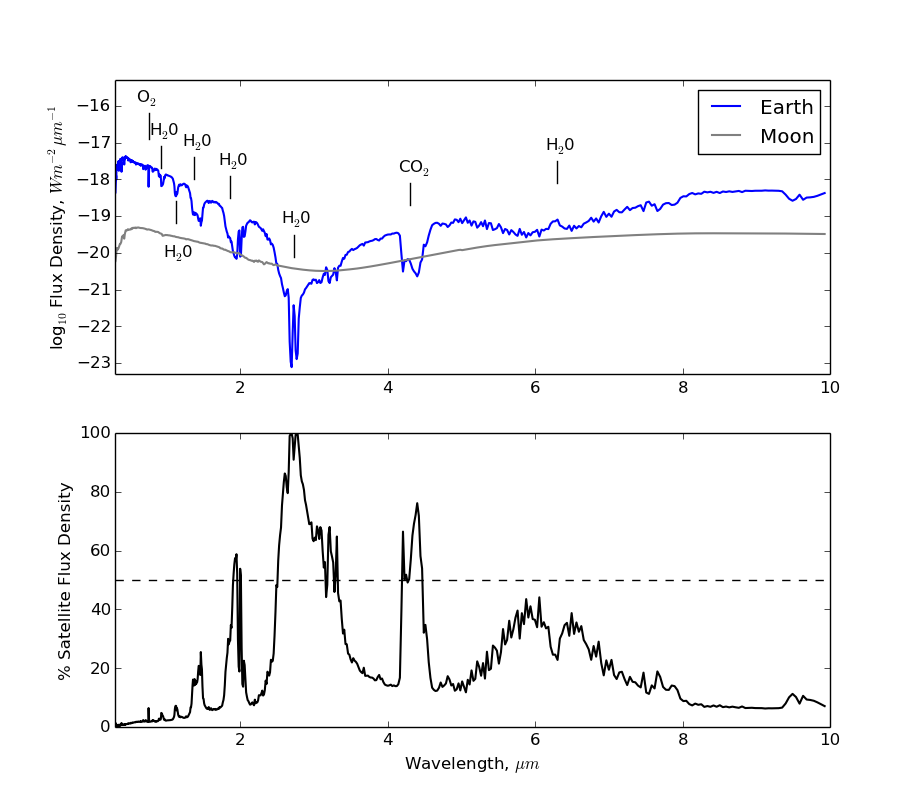}

\caption{The flux of a twin Earth-Moon system at quadrature
phase angle orbiting 
$\alpha$ Cen A as a function of wavelength in microns (top) and the contribution of flux due to the moon shown as a fraction of the total flux (bottom). The maximum fraction of the moon's flux for the Earth-Moon system occurs at $\lambda$ = 2.7 $\mu$m, contributing 99.8$\%$ to the total flux.}
\label{spectra1}
\end{figure*}

In an Earth-Moon analog system, the Moon outshines the Earth in 
the water bands around 1.96 and 2.6 - 3.0 $\mu$m, the
carbon dioxide band at 4.2 - 4.4 $\mu$m,
and it has a sizable thermal excess around 5.0 - 8.0 $\mu$m 
(Fig.\ \ref{spectra1}, top plot), although in this paper we only 
consider wavelengths out to 3 micron. The maximum fraction of the 
Moon's flux for the Earth-Moon system occurs at $\lambda$ = 
2.69 $\mu$m, contributing 99.8$\%$ to the total flux when viewed 
with spectral resolving power R = 100 (Fig.\ \ref{spectra1}, bottom plot). 

\begin{figure*}
\centering
\includegraphics[width=15cm]{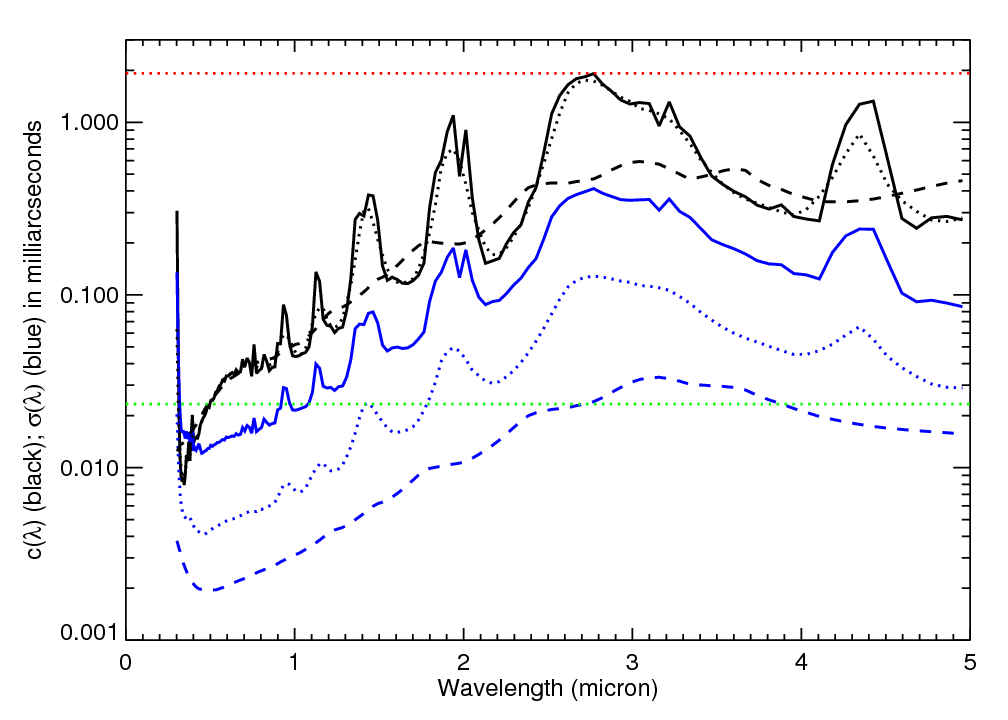}
\caption{The centroid offset, $c(\lambda)$, and uncertainty,
$\sigma(\lambda)$ for an Earth-Moon twin at
the distance of 1.34 pc ($\alpha$Cen; see Table \ref{table01}
for simulation parameters).  The reference position is
the centroid of the planet ($c(\lambda)=0$ mas), while the moon is at
$c(\lambda)=1.9$ mas (red dotted line) and the center of mass is at
0.023 mas (green dotted line). Black lines are $\bar c(\lambda)$, 
while blue are $\sigma(\lambda)$.  Solid, dotted, and dashed
black/blue lines correspond to $R=100,10,1$.}
\label{signal_noise_acena}
\end{figure*}

To optimize the spectroastrometric SNR
requires choosing two wavelengths and resolutions at which the 
difference in centroid is large, but the noise is also small.
At lower spectral resolution, the centroid offset, $\bar c(\lambda)$, 
becomes smoother so that the molecular absorption bands, at which the Moon
dominates the flux, are averaged with wavelengths where
the Earth dominates (Fig.\ \ref{signal_noise_acena}).  The noise 
decreases at lower spectral resolution due to the larger number 
of photons in each band, which
compensates for the smoothing of the centroid offset.  Consequently
lower spectral resolution generally gives a higher signal-to-noise of
detection.

\begin{figure*}
\centering

\includegraphics[width=18cm]{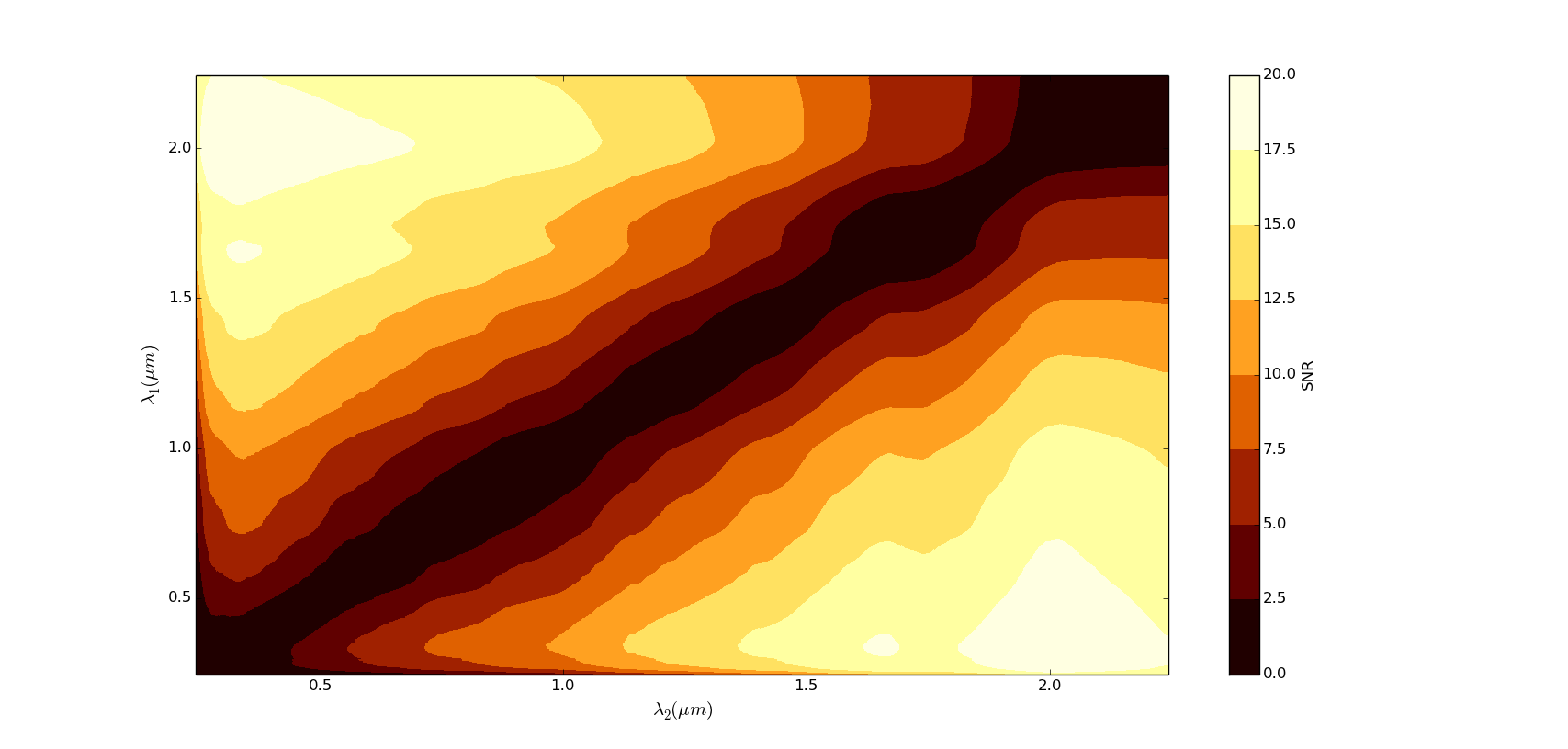}
\includegraphics[width=18cm]{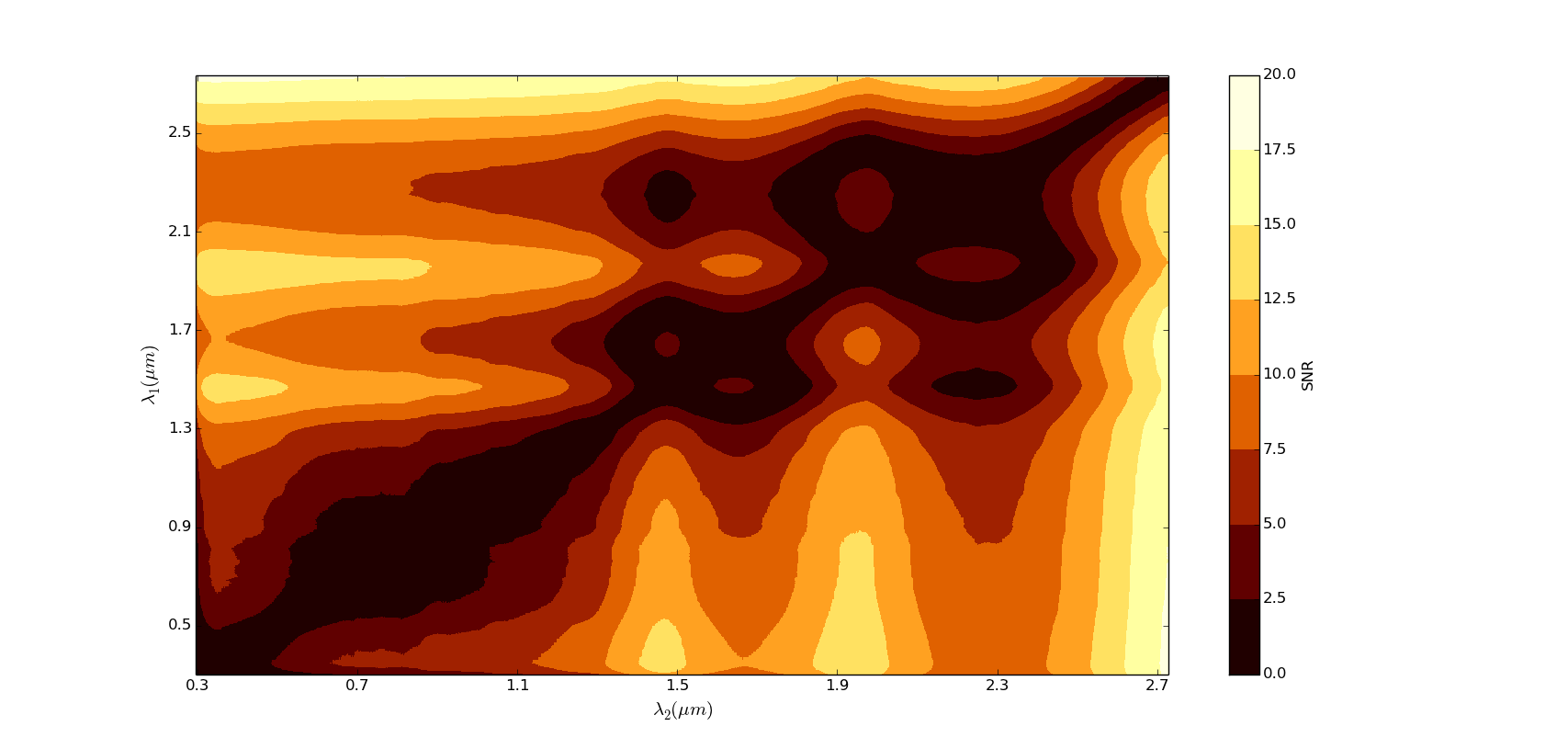}
\caption{The spectroastrometric SNR of the Earth-Moon system shown for a combination of wavelengths at R = 1.5 (top) and R = 5 (bottom). Any point on the plots corresponds to a pair of wavelength bands marked on the x and y axes. The color contours indicate the level of SNR achieved when the centroids of those two wavelength bands are compared. We assume a SNR of 5$-\sigma$ or greater would yield a detection. All calculations were done for a model system analogous to our Earth and Moon located 1.34 parsecs away in the habitable zone of $\alpha$ Cen A, with an exposure time lasting 24 hours, and a 12-m telescope diameter.}
\label{em SNR}
\end{figure*}

Our calculated SNR results for the Earth-Moon system at 1.34 pc 
exceed the 5$-\sigma$ detection threshold for the parameters 
listed in Table \ref{table01} for a wide range
of wavelengths and resolutions (Table \ref{table03}, Fig.\ \ref{em SNR}).
The optimal wavelengths cluster near $0.35$ micron and $2.7$ micron where
the planet and the moon dominate the flux, respectively.

\subsubsection{Earth-like Exomoon Orbiting a Jovian}
\label{earth jovian results}

\begin{figure*}
\centering
\includegraphics[width=14cm]{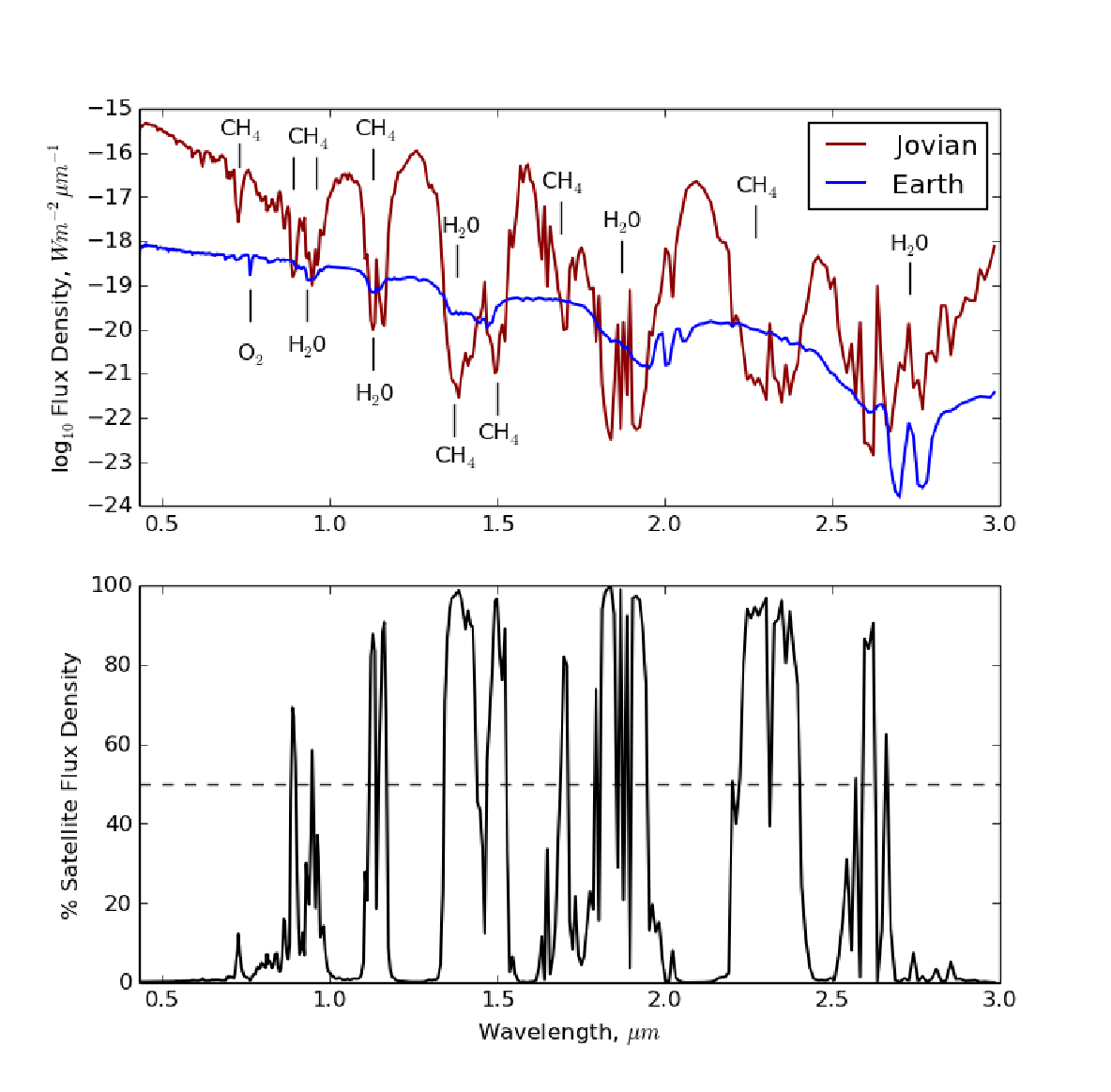}
\caption{\label{Figure2}The flux of the Earth-like moon and the warm Jupiter at quadrature phase angle as a function of wavelength in microns (top) and the contribution of flux due to the moon shown as a fraction of the total flux (bottom). The maximum fraction of the moon's flux for the Jovian-Earth system occurs at $\lambda$ = 1.83 $\mu$m, contributing 99.1$\%$ to the total flux.}
\label{spectra2}
\end{figure*}

In a Jovian-Earth system, the Earth-like moon outshines the Jovian in the 
NIR methane absorption bands, shown in Figure \ref{spectra2}. The maximum 
fraction of the moon's flux for the Jovian-Earth system occurs at $\lambda$ = 1.83 $\mu$m, 
contributing 99.1$\%$ to the total flux, although the use of this band in 
the spectroastrometric signal does not produce the highest SNR due to its higher noise;  a higher SNR is achieved in the methane band
near 1.4 $\mu$m, albeit at slightly smaller centroid offset.

\begin{figure*}
\centering
\includegraphics[width=19cm]{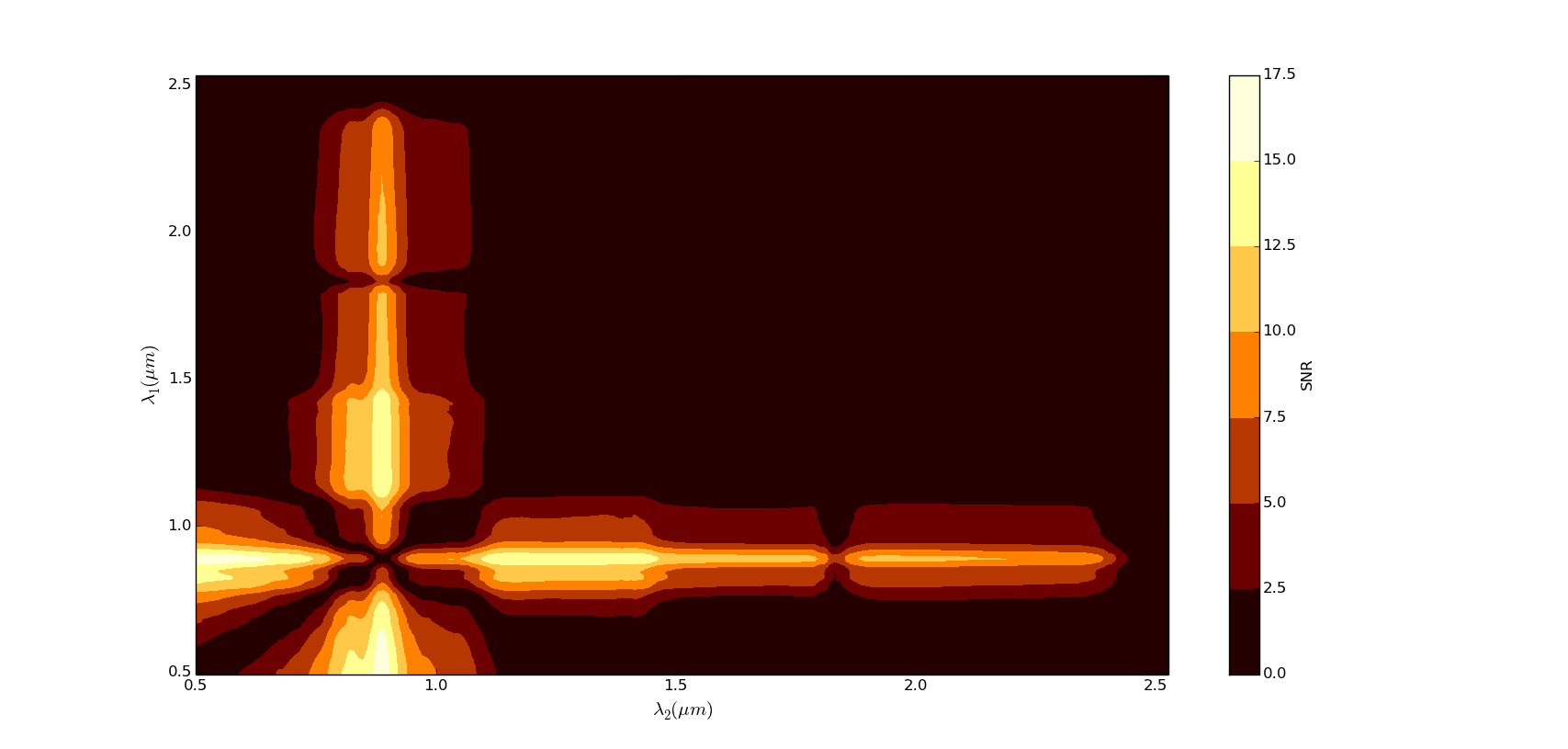}
\includegraphics[width=19cm]{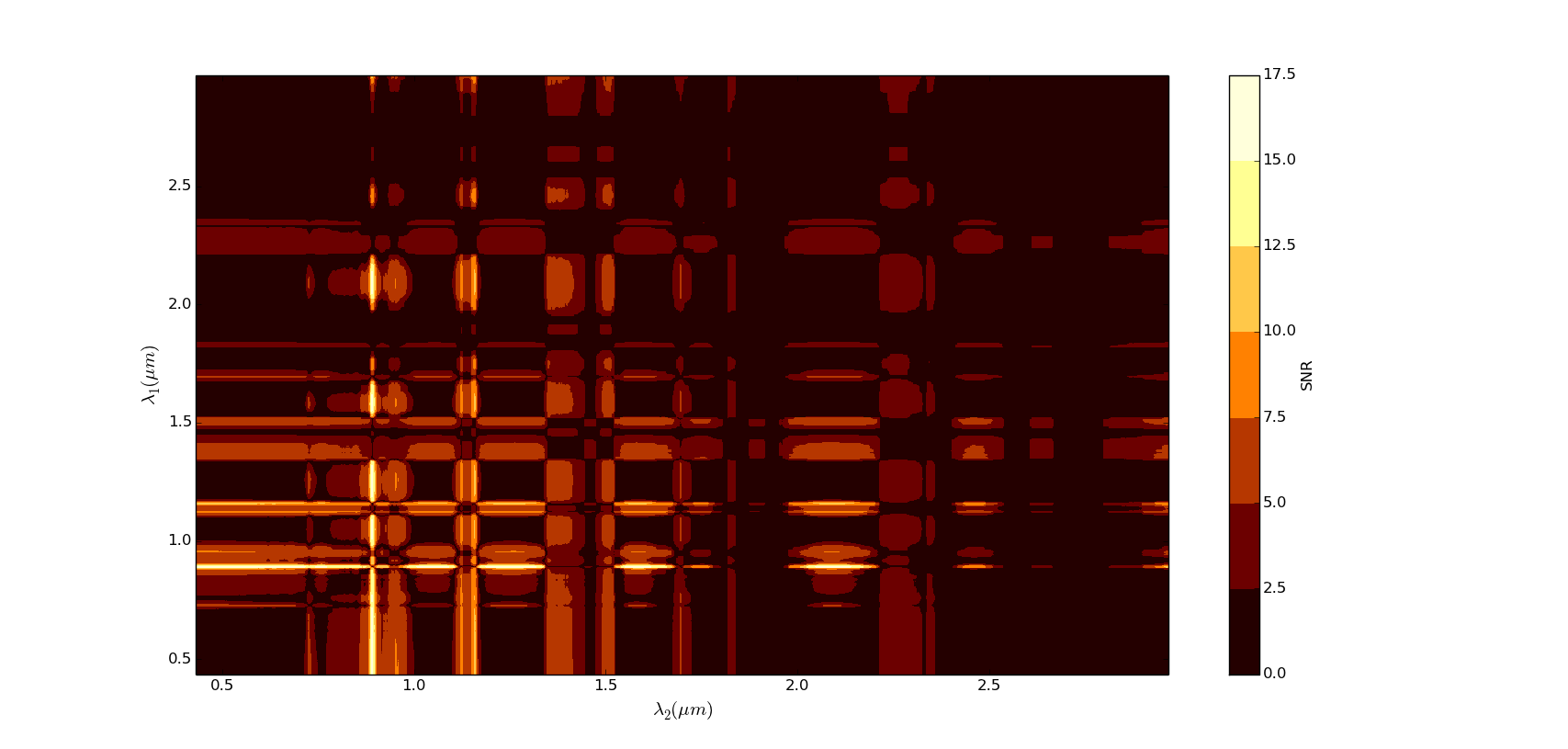}
\caption{\label{je_snr}The spectroastrometric SNR of the Jovian-Earth system shown for a combination of wavelengths at R = 4 (top) and R = 50 (bottom). Any point on the plots corresponds to a pair of wavelength bands marked on the x and y axes. The color contours indicate the level of SNR achieved when the centroids of those two wavelength bands are compared. We assume that a SNR at or above 5$-\sigma$ could yield a detection. All calculations were done for a model Jovian-Earth system located 10 parsecs away, orbiting at 1 AU around a G2V star, with an exposure time lasting 24 hours, and a 12-m telescope diameter.}
\end{figure*}

\begin{figure*}
\centering
\includegraphics[width=19cm]{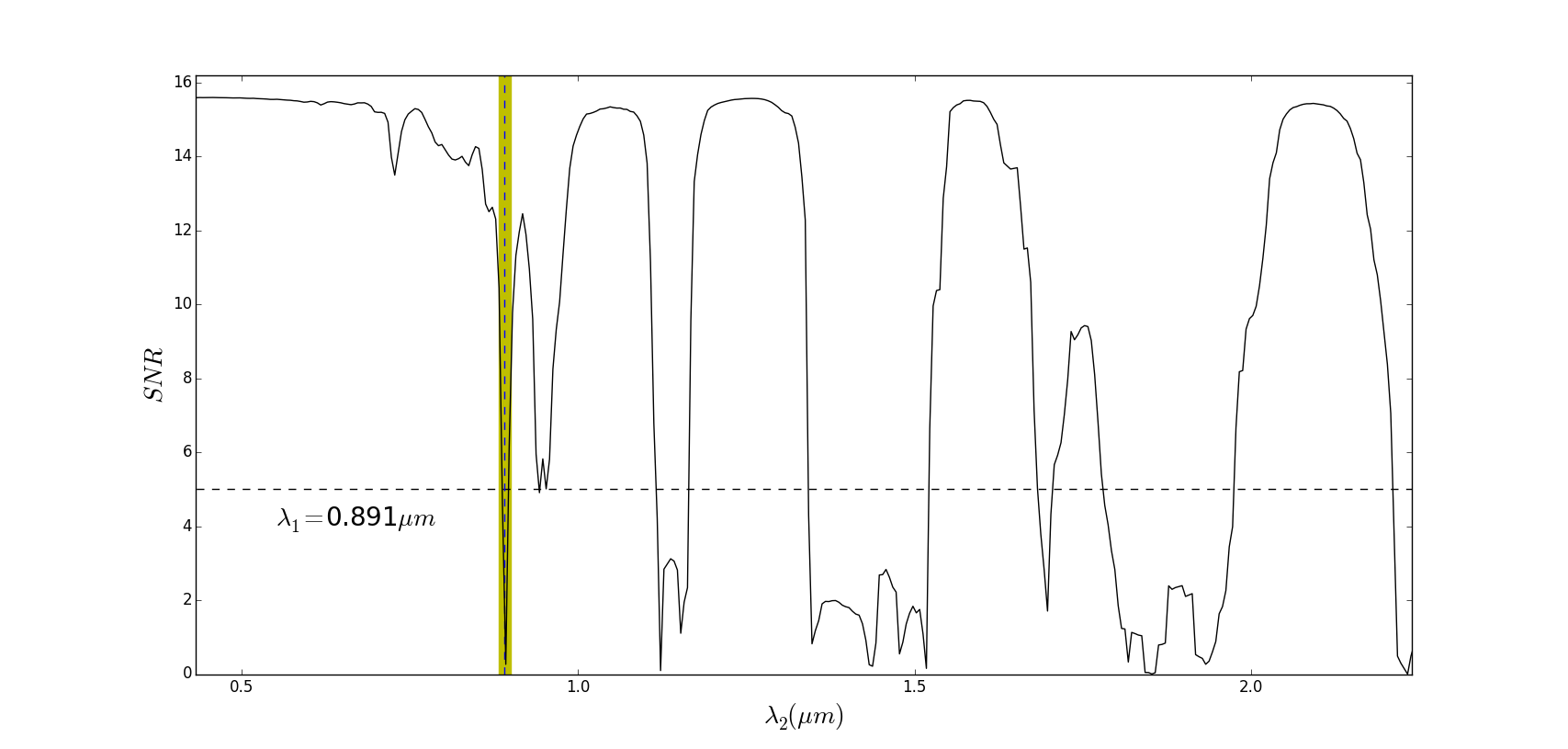}
\caption{\label{je_methane band} Methane absorption in the Jovian atmosphere leads to the moon-dominated band with a central wavelength of 0.891 $\mu m$ for R = 50 (marked for reference on figure by blue dashed vertical line and shaded green region). Fixing the centroid from this band as one of our comparisons, the plot shows the SNR (plotted as a black line) for each other band with resolving power R = 50 where signal is measured relative to that fixed 0.891 $\mu m$ band.}
\end{figure*}

The Jovian-Earth system at 10 pc produces maximum SNRs above 5$-\sigma$ for the parameters 
in Table \ref{table01}, assuming spectral resolutions between R = 2 to R = 500, and 
wavelength pairs drawn from 0.430 - 0.571 $\mu$m and 0.858 - 0.947 $\mu$m. Table \ref{table04} 
lists the optimum wavelength pairs for a selection of spectral resolutions.  To illustrate 
how the SNR changes as a function of spectral resolution, Figure \ref{je_snr} displays 
the SNR for a range of combinations of wavelengths for two different spectral resolutions. 
Note that in this model system several wavelength pairs would achieve a SNR well above
the detection threshold of 5$-\sigma$, although the optimum resolution will ultimately 
depend upon the instrument design.  Figure \ref{je_methane band} shows a slice through this plot with the choice of one
wave band centered at 0.89 $\mu$m;  this shows that a broad range of comparison wavelengths could be used.

\subsection{Characterization of an Earth-Moon system around $\alpha$ Cen A}
\label{earth_moon_characterization}

Once a planet-moon system is identified with spectroastrometry,
more telescope time will be expended on a habitable-zone system with
astrobiological interest in an attempt to
characterize the moon-planet system more fully.  If spectroastrometric
measurements can be made simultaneous to spectroscopic measurements,
then the time will concurrently be used for high precision spectroscopy 
to characterize the atmosphere and look for molecular signatures, as well
as to more precisely characterize the spectroastrometric signal. 

In particular, the spectroastrometric offset can be used to separate
the individual spectra of the planet and moon, to measure the orbit
of the moon, and to measure the mass of the planet.  As a concrete example, we
consider an Earth-Moon twin system observed for a total exposure time of one
month.  The motion of the moon about the planet will cause the centroid to
vary with time.  In the case of a circular, face-on system, the angular
motion of the centroid can be followed with time to trace out the (nearly) Keplerian
orbit of the moon about the planet.  Once an orbit is completed, the angular
motion can then be removed to give the centroid offset as a function
of time;  then, as a function of wavelength this can be binned to give
the scalar centroid offset.

We ran a Monte Carlo simulation (\S \ref{monte_carlo}) of 
an Earth-Moon twin orbiting $\alpha$ Cen A, which is the most favorable
case for detection and characterization of a moon with spectroastrometry; the proximity of this star favors many modes of 
exoplanet detection \citep{Eggl2013}.
We assume that the planet/moon instellation is the same as Earth's,
which puts the semi-major axis at 1.23 AU; planet orbits can be stable for long timescales 
at this distance given the right properties \citep{AndradeInes2014}.  
We ignore the effect of the companion star ($\alpha$ Cen B) on the observation; however,
this may contribute a potential source of noise.  Since the effective
temperature of the star ($T_{eff}=5790$ K) is slightly hotter than 
the Sun, we multiply our simulated spectra by the ratio of Planck functions
at the two temperatures, divided by the ratio of the temperatures to
the fourth power (to maintain the same incident flux).

\begin{figure*}
\centering
\includegraphics[width=15cm]{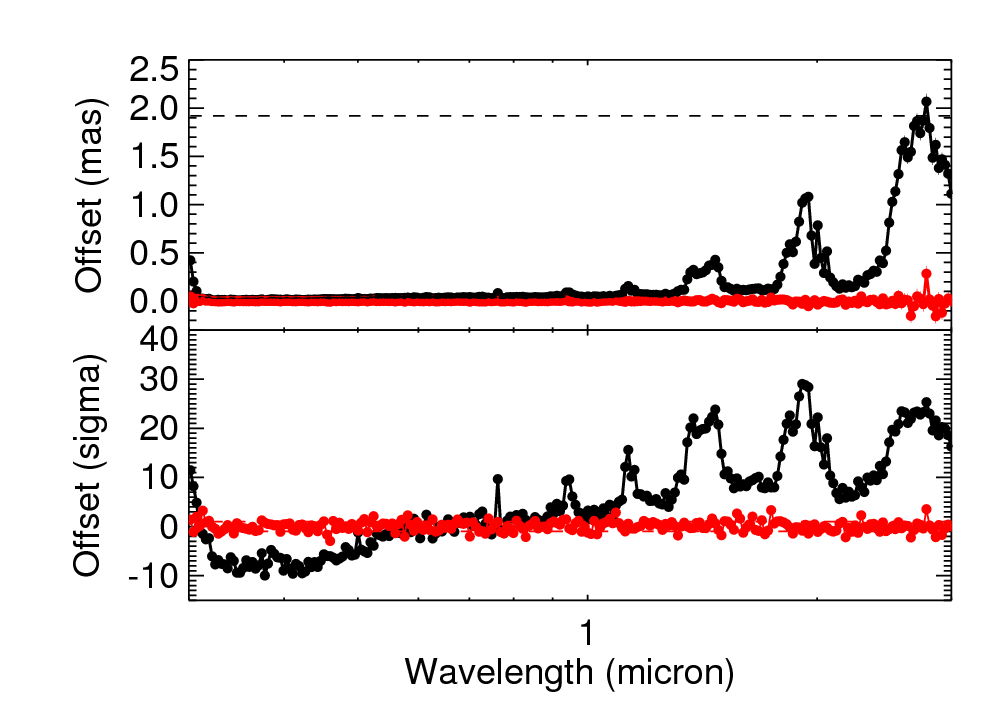}
\caption{(Top) Centroid offset, $c(\lambda)$, versus
wavelength for a simulated observation of an Earth-Moon twin
at 1.34 pc.  The orbital motion of the system has been removed
(it is assumed to be face-on).  The black line is along the
axis connecting the Earth and Moon; the red line is perpendicular.
The zero point is centered on the Earth. (Bottom) Significance
of the centroid offset relative to the mean centroid in the
directions parallel (black) and perpendicular (red) to the
line connecting the Earth and Moon.}
\label{centroid_offset}
\end{figure*}

In Figure \ref{centroid_offset}, we used the Monte Carlo simulations of
a 28-day direct imaging observation of the Earth-Moon analog system (orbiting
$\alpha$ Cen A) to measure the centroid
offset, assuming that the motion is face-on, circular, and that the
angular motion can be corrected for.  This shows that a SNR can be
achieved that varies with wavelength between the extremes of the planet centroid (at
zero) and the moon centroid (at 1.9 mas).  The detection is highly
significant, a total of 169$-\sigma$ (in practice additional sources of noise 
 may reduce this value).

\subsubsection{Spectral Distentanglement}

\begin{figure*}
\begin{center}
\includegraphics[width=13cm]{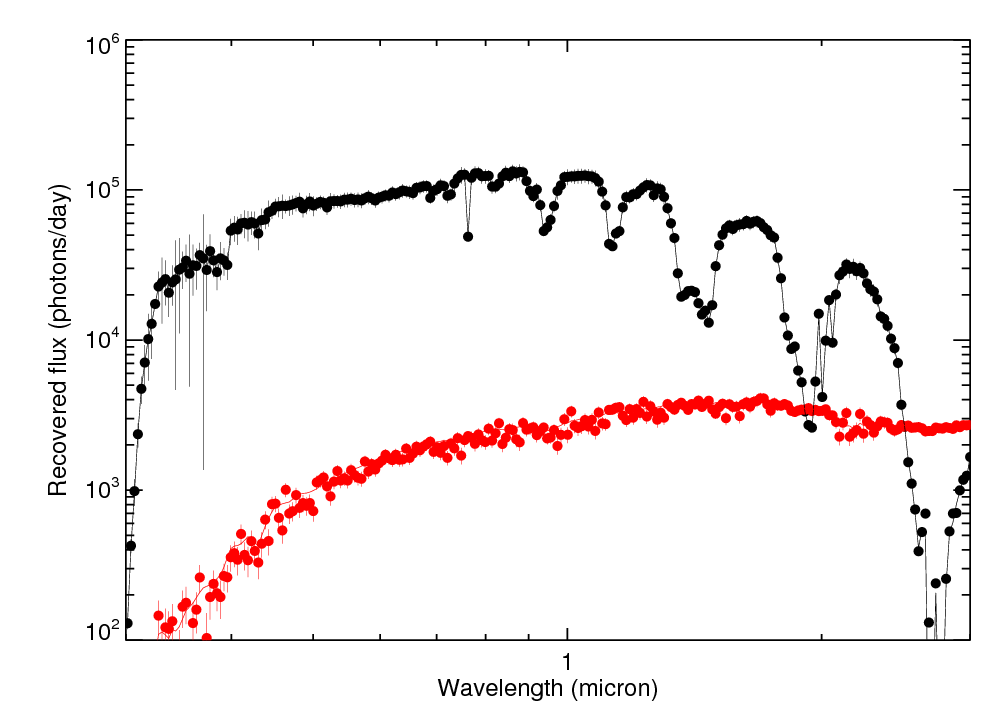}
\caption{Recovered spectra of the Earth (black) and Moon (red)
using the spectroastrometric offset.  The lines show the
input spectra; the dots show the recovered spectra.}\label{recovered_spectra}
\end{center}
\end{figure*}

We can then use this measurement to approximately separate the 
spectra of the
planet and moon as follows.  If we assume that the extremes of
the centroid motion with wavelength correspond with the positions of the planet and
the moon, then we can identify the two wavelengths at these
extremes, $\lambda_1$ and $\lambda_2$ respectively, and compute the centroid
offset signal, $S_{max} = S(\lambda_1,\lambda_2)$.  We can then
divide the centroid offset relative to the planet-dominant wavelength, $S(\lambda,\lambda_2)$,
by the maximum offset, $S_{max}$.  Then, since the center of light varies between the 
planet and the moon, the fraction of flux from the moon is simply
\begin{equation}
f_{moon}=S(\lambda,\lambda_2)/S_{max},
\end{equation}
and the planet is 
\begin{equation}
f_{planet}=1-S(\lambda,\lambda_2)/S_{max}.
\end{equation}  
We can then multiply
this by the total flux, $F_{tot}$, to recover the spectra of the individual
bodies, $F_{moon}=f_{moon}F_{tot}$ and $F_{planet}=f_{planet}F_{tot}$.  An example of 
this based on the simulated 28-day observation of the Earth-Moon twin orbiting $\alpha$ Cen A (at $d=$ 1.34 pc) 
with a 12-meter telescope and 300 wavelength bins from 0.3 to 
3 micron ($R=130$) is shown in Figure \ref{recovered_spectra}.  
The recovered spectra of the bodies match the input spectra,
and this approach allows `cleaning' the spectrum of the planet from
the contribution from the moon.  This could enable spectral
retrieval to be carried out on the individual bodies, constraining 
their atmospheric properties and compositions with further modeling.
 
If it cannot be determined whether the moon-centered wavelength
is truly dominated by the moon, then another approach is to
estimate the semi-major axis from the mass measured with
another technique, such as reflex motion of the star measured
with astrometry or radial velocity.  Then the predicted semi-major
axis can be computed from the period of the spectroastrometric
variation, and then this may be used to disentangle the spectra
assuming one wavelength is dominated by the planet.

\subsubsection{Planet Mass and Orbit Characterization}

\begin{figure*}
\centering
\includegraphics[width=12cm]{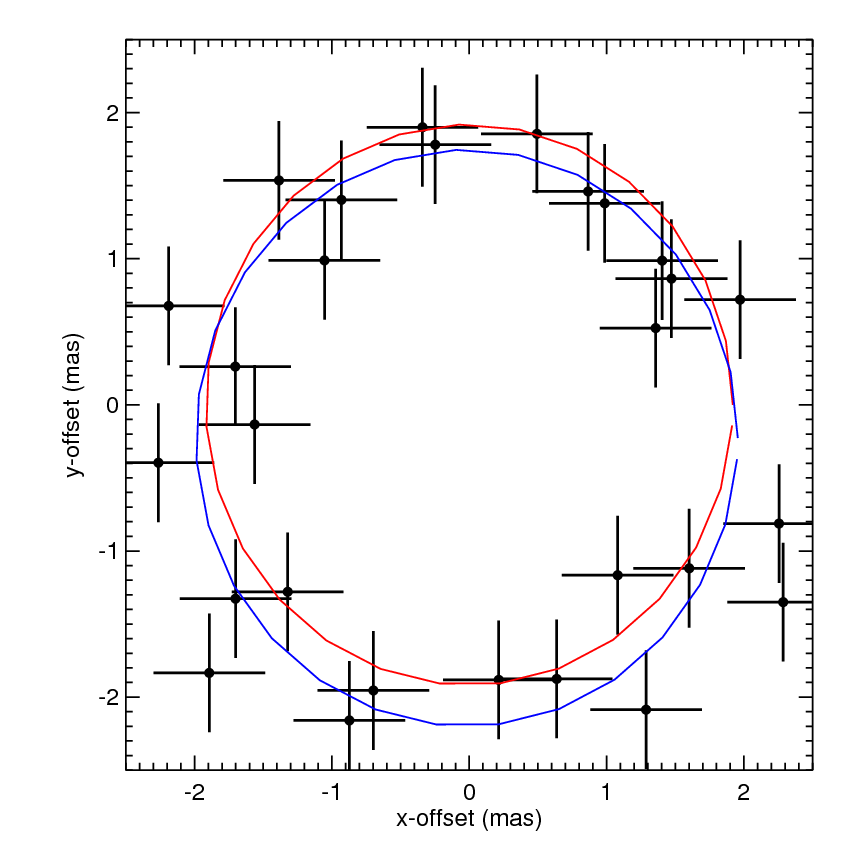}
\caption{Simulated spectroastrometric orbit of the Moon-Earth
system at 1.34 pc, which is the difference between
the $\lambda = 2.76$ and $0.35$ micron with $R=130$.  The red curve shows the input positions,
the crosses show the measured positions at 1-day intervals,
while the blue shows the best-fit ellipse for
one Monte Carlo realization of the observations.}
\label{orbit}
\end{figure*}

The time dependence of the spectroastrometric signal can be used
to measure the mass of the planet.  Based on the simulation of a
28-day signal for a face-on Earth-Moon system at 1.34 pc, we chose
the wavelength bins dominated by the Earth and Moon, respectively
at 0.35 and 2.75 micron with $R=130$.  We then measured the difference 
between the centroids at these wavelengths with time assuming 1-day exposures;
Figure \ref{orbit} shows a comparison of the input positions, the
measured positions, and the best-fit circular orbit (this can
be easily extended to an elliptical or edge-on orbit, but we
picked this face-on system as an example).  We ran 100 realizations
of the observations, and we found $M = 1.03 \pm 0.12 M_\oplus$
(the measured mass includes the mass of the moon).
This relies on having two wavelengths at which the moon and the
planet dominate the flux;  it may be possible to improve the
precision with modeling of the data, which we leave to future
investigation.

\subsection{Distance and Telescope Size Dependence}
\label{telescope parameters}

\begin{figure*}
\centering
\includegraphics[width=14cm]{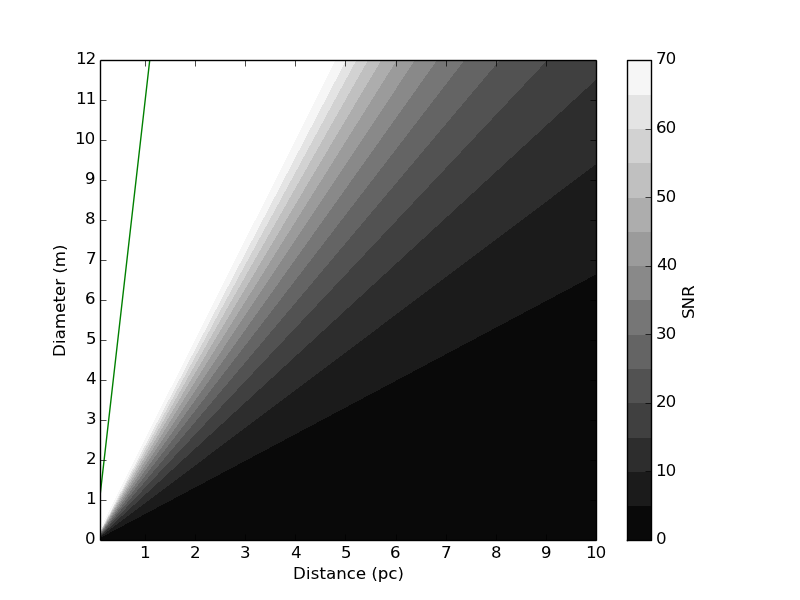}
\caption{How the SNR scales with telescope diameter and distance from the observer for the Jovian-Earth system observed between 0.45 $\mu$m and 0.89 $\mu$m at R = 50. The white region shows SNR $\geq$ 70$-\sigma$, and the grayscale shows varying SNR strengths, where the black region shows a non-detection ($<$5$-\sigma$). The green line indicates the telescope diameter needed to directly resolve the Earth-like exomoon from the Jovian planet at 0.89 $\mu$m as a function of distance, in which case spectroastrometry would not be the primary method to detect an exomoon.}
\label{snr_distance}
\end{figure*}

\begin{figure*}
\centering
\includegraphics[width=14cm]{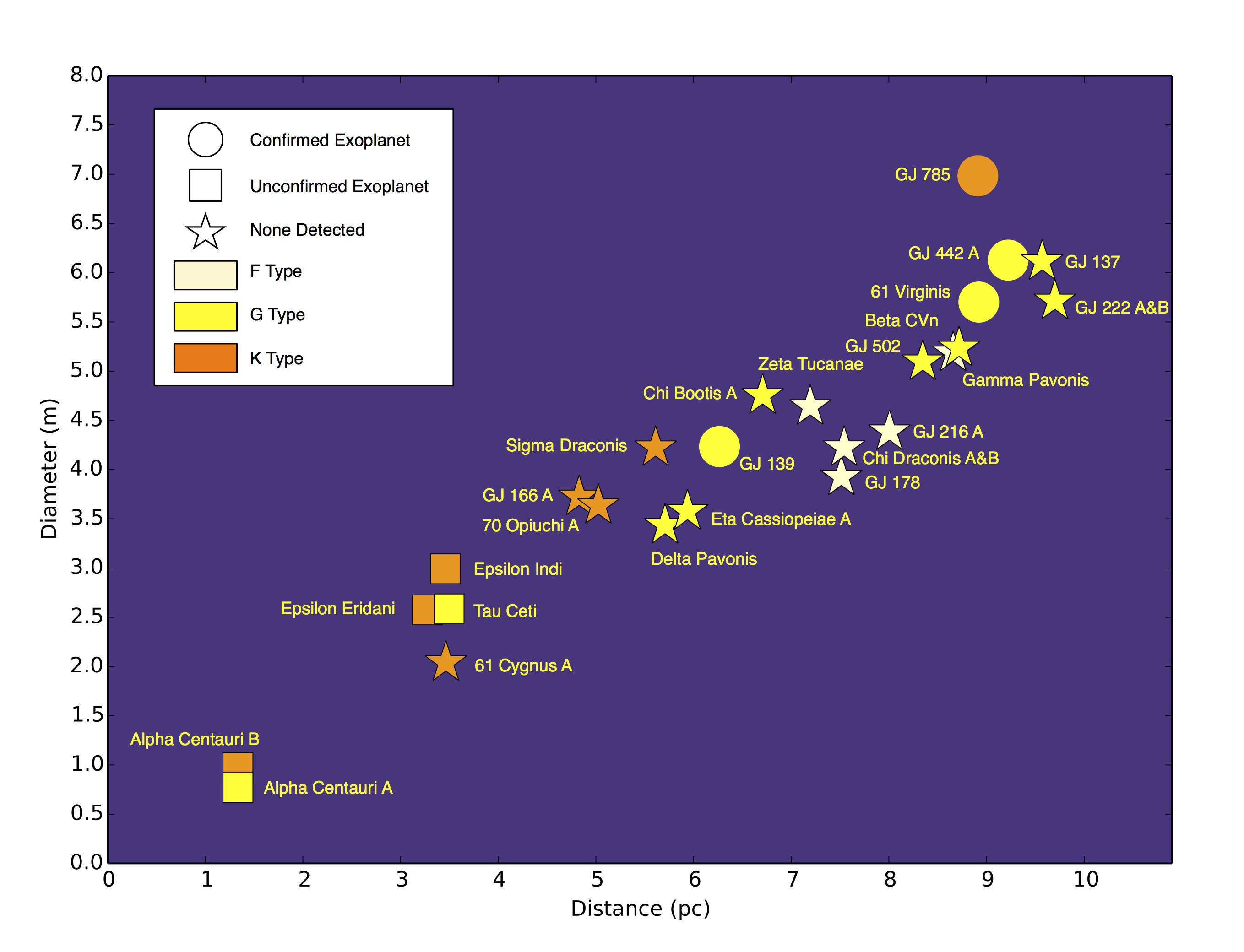}
\caption{Minimum telescope diameters needed to obtain a 5$-\sigma$ spectroastrometric detection between 0.45 $\mu m$ and 0.89 $\mu m$ of an Earth-like exomoon orbiting a warm Jupiter in the approximate habitable zones of the systems labeled above, displayed at their respective distances. The different colors represent the spectral types F (pale yellow), G (yellow), and K (orange). The circles represent the systems with confirmed exoplanets, the squares for systems with unconfirmed exoplanets, and the stars for systems in which exoplanets have yet to be detected.}
\label{nearby_stars}
\end{figure*}

The SNR's simultaneous dependence on distance and telescope diameter for the Jovian-Earth system is shown in Figure \ref{snr_distance}. Using the wavelength pair $\lambda_{1}$ = 0.45 $\mu$m and $\lambda_{2}$ = 0.89 $\mu$m at R = 50 and $t_{obs}$ = 24 hrs, a 6.7-m telescope would be sufficient to detect an Earth-like moon orbiting a Jovian around a Sun-like star at 10 pc using spectroastrometry with a confidence of 5$-\sigma$. According to the Gliese Catalog of Nearby Stars, there are about 20 main sequence F, G, and K type stars within 10 parsecs that could be used as potential targets. These systems and the diameters needed to detect such a planet-moon system with a confidence of 5$-\sigma$ may be seen in Figure \ref{nearby_stars}.  Since the effective temperature of each star differs from  
the Sun, we multiply our simulated spectra by the ratio of Planck functions
at the two temperatures, divided by the ratio of the temperatures to
the fourth power (to maintain the same incident flux, which is assumed to
be the Solar constant).  This assumes that the albedos of the planets
remain the same for planets orbiting stars of different spectral types, and that the 
stellar spectral features differ slightly;  both assumptions are approximate, 
but we expect will have a small effect on our results compared to, for
example, planets with different atmospheric compositions that what
we have simulated.

The SNR's dependence on distance from the observer arises from both the signal's inverse relationship to the distance and the flux's contribution to the noise, where the total flux $F_\lambda$ is inversely proportional to the square of the distance \textit{d}.  Equation \ref{snr_scaling_equation}
shows that the SNR scales as $d^{-2}$, where one power comes from
the decreasing angular separation of the moon and planet, and one
power comes from the noise caused by decreasing flux of the moon.

The telescope diameter affects the width of the point spread function as $D_{tele}^{-1}$, as well as the total number of photons incident on the detector in a given exposure time $t_{obs}$ as $D_{tele}^2$, giving a
scaling of the SNR as $D_{tele}^2$, as shown in equation \ref{snr_scaling_equation}.  Given the overall dependence of
SNR as $(D_{tele}/d)^2$, the volume which can be surveyed to a 
fixed SNR scales as $d^3 \propto D_{tele}^3$ (holding other parameters fixed).
If the diameter of the telescope is decreased, then the observing
time needs to increase as $t_{obs} \propto D_{tele}^4$ to obtain an equivalent SNR;
thus, either fewer systems can be surveyed due to the longer observing
time, or the observing time will increase until systematic errors dominate the noise.

We have focused on habitable-zone planet/moon systems around G stars in this
paper, but moons should be searched for around planets at a range
of star-planet/moon separations.  More distant planet-moon pairs
will have lower intensity due to the decreased flux they receive
from the star:  $F_\lambda \propto a^{-2}$, where $a$ is the semi-major
axis of the orbit of the planet and moon about the star.  
However, their Hill spheres will expand due to
their larger separation from the star, so if the moon's orbital
distance is a fixed fraction of the Hill sphere, then the spectroastrometric
signal will increase $\propto a$ due to the wider moon-planet 
separation.  These
two effects compensate to cause an equal SNR as a function of
the distance of the planet/moon system from the host star.

\section{Discussion}

The detection of an exomoon is a long-sought discovery which could
profoundly impact our understanding of planet and satellite formation 
and evolution, including that of our own Solar System.
It is clear that moons can outshine their planets at some wavelengths
and contribute to apparent variability of a planet
\citep{2004AsBio...4..400W,2009AsBio...9..269M,2011ApJ...741...51R};  
in practice, though, it may be difficult to prove that spectroscopic 
or time-dependent signatures are in fact due to a moon.  Since a moon 
should be spatially separated from a planet, and the separation should 
follow a (nearly) Keplerian orbit as a function of time, spectroastrometry 
gives a means of detection that is more definitive, and would allow 
measurement of the properties of the planet-moon system.  

If the separation of the planet and moon is similar to the 
width of the instrumental point-spread function (PSF), then one 
can simply resolve the planet and moon.  For example, a favorable 
configuration of a satellite orbiting at 0.25 of
the Hill radius of a Jovian-mass planet orbiting at $\approx 1$ AU around 
a star at 3 pc will have a maximum angular separation of $\approx 6$ mas.
This is about the angular resolution at 0.5 $\mu$m of an 8-meter space
telescope; in practice, the angular separation will be smaller, so most planet-moon systems will be unresolved,
hence the need for spectroastrometry. The $\alpha$ Cen A case of an 
Earth-Moon analog can {\it almost} be resolved directly:  
at 0.35 $\mu$m, a 12-meter telescope has a resolution of $\approx 2.7$ mas,
while the angular separation of the Earth-Moon analog is $1.9$ mas.
Since the Moon is much fainter than the Earth at these wavelengths
($\la$ 1\%), it may be difficult to deconvolve the light, and thus
spectroastrometry will still be favored for detection of the moon.

We have chosen 3 micron as the long-wavelength limit for these observations
based on the fact that current coronagraphic designs are considering 2 micron
as a possible wavelength cutoff, with a possible extension to 3 - 5 micron
\citep{Dalcanton2015};  longer wavelengths have the problem of a larger
inner working angle and noise due to thermal emission from the telescope.
Spectral coverage that extends to 3~$\mu$m covers the water band at 2.7~$\mu$m and the
methane band at 2.3~$\mu$m;  these bands have the advantage of being dominated
by the moons in the two cases we have considered.  In particular, for
measuring the mass of the Earth-like planet in the Earth-Moon analog
case, the 2.7~$\mu$m band is critical for being able to measure the
planet-moon semi-major axis.  At 3~$\mu$m we find that there are
$\approx 7$ KV stars out to 5 parsecs which can be resolved at
$> 2 \lambda/D_{tele}$ for $D_{tele}=12$ meters; 15 GV stars out
to 17 parsecs; and 36 FV stars out to 24 parsecs (we have not considered
stars beyond 25 parsecs).  The possibility of detection of an exomoon
in these systems depends strongly on the potential properties
of the exomoon-exoplanet systems present, including their frequency.

Another issue we have not considered
is the effect on these observations of exo-zodiacal light due to
scattering/thermal emission by dust.  This diffuse emission component could have a spectrum similar 
to that of a moon, and if concentrated in a region near the planet, may be confused with the effect of a moon.  If spread
throughout the system it would contribute to the noise of the observation,
which grows more severe at longer wavelengths due to the larger PSF, which
scales in solid angle (or area on the detector) as $\lambda^2$.  If
(exo-) zodiacal light is a significant source of noise, then shorter 
wavelengths may be favored.  We estimate \citep{agol2007} the
ratio of the flux of the exozodi flux within the planet's PSF,
$F_{EZ}$,
to flux of the host star, $F_*$, to be:
\begin{eqnarray}
\frac{F_{EZ}}{F_*} &=& \frac{d^2}{r^2} \tau_{EZ} S_{fac} \left(\frac{\lambda}{D_{tele}}\right)^2\cr
&=& 6 \times 10^{-12} d_{pc}^2 r_{AU}^{-2} \frac{\tau_{EZ}}{10^{-9}}
\left(\frac{\lambda}{3 \mu{\rm m}}\right)^2 \left(\frac{D_{tele}}{12 {\rm m}}\right)^{-2},
\end{eqnarray}
where $\tau_{EZ}$ (units sr$^{-1}$) relates the incident stellar flux to the exozodi
surface brightness, and $r$ is the planet-star separation.  This
is less than the Moon flux for Solar-System like
exo-zodiacal light, but it can increase with distance, wavelength, 
or dust surface density, and thus may play a factor for more distant,
young systems.
Older systems may be largely void of dust (like our 
Solar System).

The precise choice of the wave bands used for detection will depend on multiple
factors rather than just optimizing the signal-to-noise ratio.  For example,
if a telescope has a short or long-wavelength cutoff that differs
from what we assumed, this may require a different choice of band.
If a narrow wavelength range is required, then it may be necessary
to choose more closely spaced wavelengths than what we have chosen.
As an example, consider the fiducial Earth-Moon twin system with 
parameters from Table \ref{table01}.  We tried all pairs of wavelengths from
0.3 - 3 micron, and resolutions from R=1-10 (in steps of 1) which
give SNR $>10$ for this simulated system (for a 12-meter telescope
and 1-day observation).  From these, we found
wave bands that give the largest spectroastrometric signal,
$S(\lambda_1,\lambda_2)$:  we find that $\lambda_1,\lambda_2 = 0.335, 2.73$
micron and R$=10$, gives a centroid offset that is 90\% of 
the Moon-Earth sky separation at quadrature (SNR = 14).  Higher resolutions 
can approach 100\% (Fig.\ \ref{signal_noise_acena}), but at lower SNR; this is
what was required for the characterization of the system (\S \ref{earth_moon_characterization}).

Another consideration is the separation in wavelength:
if the instrumental systematic errors of the centroid measurements grow with the separation
in wavelength, then it may be advantageous to choose two bands that
are closely spaced in wavelength.  Amongst the SNR $>10$ band pairs
for the Earth-Moon system at 1.34 pc, we find $\lambda_1,\lambda_2 = 2.3,2.7$
micron (R=10) gives a centroid offset that is 75\% of
the sky separation.  The precise choice of bands and resolutions
will depend strongly on the planet properties and stellar distance,
as well as the properties of the coronagraphic telescope, so these
examples are merely for illustration.

With the detection of a moon with spectroastrometry, there are several
applications of the detection which may be used for characterizing the
planet-moon system.  A spectroastrometric detection will confirm that a
spectroscopic moon signal is truly due to an exomoon, and not, for example,
due to varying temperature or gas mixing ratio profiles within the planet's 
atmosphere; it will
also confirm that the planet is in fact a planet.  The variation of the
spectroastrometric signal with time will follow the orbit of the moon
about the planet, giving the period of the moon's orbit when monitored with
time.  As shown above (\S \ref{earth_moon_characterization}), using
Kepler's law, the period and 
semi-major axis allow the determination of the mass of the planet plus moon;
this could enable
the mass measurement of an Earth-Moon twin at 1.34 pc with a 12-meter
telescope at 0.35 and 2.76 micron. This will also
potentially allow the determination of the semi-major
axis, orbital inclination, and eccentricity of the moon, if the signal-to-noise is
sufficient.  These applications require that wavelengths can be 
identified at which the planet
and moon each dominate the flux, although this may be difficult to prove
in practice as the spectroastrometric signal is degenerate
between the moon's radius, $R_m$, and semi-major axis, $a_{mp}$, $S \propto R_m^2 a_{mp}$.

In addition to the spectroastrometric
signal, which yields the difference of the center-of-light between the
moon and the planet, if the absolute astrometric signal of the planet can be
measured, then it might be possible to also measure the mass of the
moon due to the reflex motion of the planet about the center of mass with
its moon. However, this signal, which is of order the size
of the Earth for the Earth-Moon system, may be swamped by the contribution
of the Moon's signal to the centroid, which would need to be removed
precisely to measure the offset of the Earth from the center of mass
with the Moon;  this will require identification of the wavelength
at which the Moon-Earth flux ratio is less than the Moon-Earth
mass ratio, such as near $0.35$ micron (Fig.\ \ref{signal_noise_acena}).
In addition, the center-of-mass offset of the planet's light is in
the opposite direction of the of the center-of-light offset (relative
to the center of mass) at wavelengths where the moon
has a larger flux fraction.  There is a wavelength at which the 
two offsets cancel, at which point the center of light coincides with 
the center-of-mass;  for the Earth-Moon system at quadrature this occurs 
at 0.493 $\mu$m (Figure \ref{signal_noise_acena}).
The astrometric signal will also be affected by the non-uniform
illumination and non-uniform albedo (or thermal emission) of the Earth, 
which will vary with orbital phase and rotation of the planet, which 
should be accounted for if the astrometric precision
is sufficiently high.  This could, in fact, provide another opportunity
to detect spatial variations in the surface of the planet, and potentially
reveal the full obliquity of the planet, recirculation of heat (which
would be affected by the atmospheric pressure and opacity of the planet),
and surface markings (or persistent cloud features).
In addition, perturbations by other planets in
the system can affect the motion of the center-of-mass;  for the
Earth-Moon system, these perturbations also have an amplitude that is
comparable to the radius of the Earth.  Despite these
complications, this measurement could also be attempted with a
coronagraphic telescope capable of precision astrometry of planets,
which could be measured with respect to the position of the host star.
A diffractive element introduced into a high-contrast imaging
coronagraph can yield the astrometric position of the star, and thus 
allow the precise measurement of the planet's orbit with respect to the star
\citep{Guyon2013}.

A spectroastrometric detection of an exomoon can potentially result in
a detection of eclipses/transits/mutual events. The orbit of the moon 
about the planet measured from spectroastrometry may allow one to forecast 
whether and when a moon/planet will transit.   Knowing whether the events will occur
and observing at the forecast times gives a significant advantage over 
trying to initially detect an exomoon with eclipses/transits/mutual 
events for which continuous observations are required and a successful 
detection depends on the unknown geometry of the orbits 
\citep{2007A&A...464.1133C}.  There are four possibilities of the system 
geometry to consider \citep{Schneider2015}:  1) if the
moon's orbit has a small inclination with respect to the planet's orbit
(which can be measured from its motion about the star with time), then
the moon can regularly pass between the star and the planet, causing 
a shadow to be cast on the planet, which may be visible by a dimming of 
the planet at 
the predicted time of the stellar eclipse; 2) the passage of the moon 
into the planetary shadow (with respect to the star)
can also yield a dimming due to the lunar eclipse; 3) if the inclination 
of the moon relative to our
line of sight is nearly edge-on, then one might observe a transit of the
planet by the moon, causing a dimming at the predicted time \citep{Livengood2011}; 4) likewise,
the moon can pass behind the planet.  It is possible for cases 1/2 and 3/4
to occur simultaneously (given a fortuitous geometry), causing an even
larger depth. The latter two cases have a
$\approx 2$\% probability for a randomly placed observer of the Earth-Moon
system, which is about four times larger than the probability of viewing
the Earth to transit the Sun ($\approx 0.5$\%).  Any four of these configurations,
if detected at sufficient signal-to-noise, might yield a measurement of
the radii of the planet and/or the moon (or their ratio), giving a means of determining
the density of the planet (and possibly the moon).  In addition, if the 
spectral dependence of
the transits/eclipses are measured, this might yield the detection of
spatially-resolved features on either body.   \citet{HellerAlbrecht2014}
have also studied the possibility of measuring the Rossiter-McLaughlin
effect in high-resolution spectroscopy of resolved exoplanet
being transited by an exomoon;  initial spectroastrometric detection
could be used to forecast the timing of these events.

Spectroastrometry may also enable a measurement of the spectrum of the moon.
In the case of an Earth-like planet orbiting a giant planet in the
habitable-zone,  this may be one of the best ways to characterize the
atmosphere of the planetary companion.  Once the solution for the
moon-planet orbit is found, the observations can be integrated for a longer
time, possibly allowing a spectrum to be built up of both 
bodies (\S \ref{earth_moon_characterization}).  
This
technique works best if two wave bands can be pinpointed within which
the planet dominates in one band and the moon in the other; then the
spectroastrometric signal between these two bands gives the sky separation 
of the bodies at each time. 
In the example we considered of a face-on circular orbit (\S 
\ref{earth_moon_characterization}), the planet-moon separation is unchanged with time.
For a different orbital configuration, the planet-moon separation
needs to be modeled as a function of time and wavelength with a Keplerian orbit.
Then the size of the orbit as a function of
wavelength (with respect to a reference wavelength) can be used
for the spectral decomposition.
The edge-on case will have a signal that is reduced by $\approx \sqrt{2}$.
This measurement can
possibly allow the correction of a planet's spectrum for contamination 
by a moon, avoiding a potential false-positive for disequilibrium
chemistry \citep{2014PNAS..111.6871R}.  The recovered spectra of
both bodies could constrain
their albedos and their thermal emission properties, further
constraining the atmospheric properties.

With measurements of the orbital and bulk properties of the moon-planet
system, one can potentially place constraints upon the formation of
the system, as well as its tidal evolution.  In particular, the location
of the orbit within its Hill sphere, and the eccentricity of the orbit,
will constrain tidal migration and circularization of the system, which
can potentially yield a constraint upon tidal evolution theory \citep{2014AsBio..14..798H}.

We end by mentioning some other issues to consider in future work.
We have only considered two example planet-moon compositions, and a limited
range of an (admittedly) large parameter space.  The sizes of the planet and moon,
spectral type of the host star, planet-moon separation, spectral
variability of the planet and/or moon,
semi-major axis of the stellar orbit, 
exo-zodiacal light, presence of other planets,  presence of rings (about the
planet or star),  inclination of the orbital axes, eccentricities, and more 
can affect the detection.  Most importantly, the composition and presence 
(or absence) of an atmosphere, albedo, presence of clouds or hazes, and 
other properties of the atmospheres or reflective surfaces will affect 
which bands should be observed, and at what wavelengths each body will
dominate.  Planet and/or moon variability should be considered as
well.  In the case of the Earth-Moon system, the quadrature Earth varies 
at the level of roughly ten percent over its rotation
period (one day).  This would be detectable in total flux, and so
the spectrastrometric signal could be averaged over the rotation
timescale, assuming that the two bands could be measured at the
same time.  Longer timescale variability could affect the orbital
characterization and spectral disentangling;  we leave the issue
of variability for future work.  Tidal heating of an exomoon could
enhance the spectroastrometric signal if the temperature were
increased to the point of contributing strongly in the absorbing
bands of the planet.
Multiple moons (or rings) could affect the interpretation of the 
spectroastrometric signal; we have assumed a single moon based on the
logic that the first detections will be of large moons, and large
moons might not exist stably with multiple moon companions.  The inclination 
of the moon with respect to the observer will affect the spectroastrometric 
signal:  an edge-on moon will spend some time at a small projected separation 
from the planet, and so an observation at the wrong time might miss the spectroastrometric 
signal.  This will require additional observing time to sample the (unknown)
phase of the moon sufficiently to be able to capture it at maximum elongation.  
We have only considered observations at quadrature illumination, while the orbital phase 
dependence of the spectroastrometric signal may yield additional information 
about the bodies.  Also, in computing the spectroastrometric signal, we
have ignored the motion of the moon.  This should be adequate if
the duration of the observation is much shorter than the orbital period of
the moon, but for long exposure times or short moon periods, this could
affect our signal-to-noise computations.  Once a moon is detected,
and its orbital period measured, the orbital motion can be accounted
for in modeling the spectroastrometric signal as a function of time.

The design
of the detector which would allow spectroastrometric measurements needs
to be explored.  New detector technologies that allow for the measurement
of the position and energy of a photon could potentially
yield low-resolution spectroscopic images, and the resulting data cube could
be binned in wavelength channels to optimize the detection of a moon,
and later to optimize the characterization of the spectra as a function
of wavelength.  Specifically, Microwave Kinetic Inductance Detectors, or
MKIDs, and Superconducting Tunnelling Junctions, STJs, 
have high quantum efficiency and are being multiplexed into larger arrays
\citep{Peacock1996,Day2003,Mazin2012}.  Another possible approach would be to use an
integral field unit spectrograph, such as that used on the Gemini
Planet Imager \citep{Macintosh2006}.
The wave bands that are used could in principle be
selected after the observations are made:  the smaller body tends to
dominate at wavelengths where the larger body is dimmest, so the wavebands
can be selected based on the observed molecular band features that are
found after the initial spectrum is measured.  Then the centroid offset
can be searched for by measuring on-band and off-band.
The spectroastrometric precision may be affected by the
sampling of the point-spread function as a function of wavelength;  this
should be studied further to determine what pixel/fiber size is required 
to obtain astrometric precision at each wavelength that is limited by the 
width of the PSF and number of photons detected, and not its spatial sampling.  

We have limited our exploration to a coronagraphic telescope operating
out to 3~$\mu$m.  Interferometric
approaches might allow spectroastrometry to be extended to longer wavelengths,
where additional absorption bands could be used and where the thermal emission from
the moon can be significant, such as in the TPF-I/Darwin concept operating
from 6 - 20 $\mu$m \citep{Lawson2008,Cockell2009}.  This would have great advantages
for characterizing the planet-moon system, but involves the significant technical
challenges of formation flying, baseline control, and thermal backgrounds.
In principle this detection technique could be applied to ground-based
direct imaging as well with future large coronagraphic telescopes; \cite{Schneider2015}
have mentioned the possibility of detecting the astrometric wobble
of a directly imaged giant planet due to an exomoon companion.
The precision may be limited more strongly by control of the optics
and atmosphere, but nonetheless could possibly detect large moons in
the infrared, and spectroastrometry could potentially be more sensitive
than absolute astrometry depending on the instrumental and observational
design.

One challenge of coronagraphic spectroastrometry is that precise
centroiding requires controlling sources of noise, as well as a stable PSF 
that can be measured astrometrically over a broad range of wavelengths, 
requiring stable pointing and control of the PSF shape.  These
requirements may drive  
stricter tolerances for the design of a high-contrast imaging telescope
than the requirements for spectroscopy.  For
example, the statistical uncertainty for the 28-day observation
of a 1.34-pc Earth-Moon system with a 12-meter telescope is 
$\approx 3$~$\mu$arcsec, and the typical ratio of the astrometric precision
can be as small as 0.01-0.1\% of the width of the PSF.  Thus, the
pointing, the variation of the PSF, the characterization of the PSF,
and the effects of other sources of noise need to be limited/controlled/calibrated
to a very precise level.  The telescope design
and detector calibration would need to be capable of reaching this 
sub-pixel precision in measuring the location of the centroid of the 
planet-moon PSF as it passes across the detector over time;  sub-pixel 
sensitivity variations, detector latency, charge bleeding, and light 
loss would all have to be controlled or calibrated to high precision 
to reach the photon-noise limit that we have assumed for this paper.
One mitigating factor is that the spectroastrometric signal changes
sign and direction as the moon orbits the planet, and should repeat
with time, while any systematic errors will likely have a different
time dependence.
 
\section{Conclusions}

Spectroastrometric detection presents a scientifically promising,
but technically challenging, means of detecting and studying
satellites of exoplanets, `exomoons.'  We have presented two case studies 
to illustrate the potential for detection of this effect, and
urge that studies of future coronagraphic telescope designs take
into account the spectroastrometric signal when designing the
technical requirements of telescopes and detectors.  Subsequent
to  detection, the characterization of exomoons with
spectroastrometry 
could lead to measurement of the fundamental properties of
exoplanet-moon systems, including the mass of the host planet,
and could help in pinpointing targets 
that are valuable for studies of habitability.

We end by making some general recommendations for guiding design of
future coronagraphic space telescopes with spectroastrometry
in mind:  1) the instrument
suite should include the capability of making astrometric
measurements as a function of wavelength for a range of
spectral resolutions; 2) the pointing control, PSF wing suppression,
and PSF stability and calibration
should be designed with the capability of making photon-limited
astrometric measurements over a broad range of wavelengths;
and 3) the instrument wavelength should be extended out to $\approx 3$
micron to cover the water absorption band at which the
Moon dominates over the Earth ($\approx 2.7$ micron)
and methane band at which the Earth dominates over
Jupiter ($\approx 2.3$ micron).

\section*{acknowledgments}

TR, EA and VM were partly funded by the NASA Astrobiology Institute's 
Virtual Planetary Laboratory, supported by the National Aeronautics 
and Space Administration through the NASA Astrobiology Institute under 
solicitation No.~NNH05ZDA001C. We thank Rory Barnes and referee Ren\'e 
Heller for feedback on the submitted version of this paper.
TR gratefully acknowledges support from 
an appointment to the NASA Postdoctoral Program at NASA Ames Research 
Center, administered by Oak Ridge Affiliated Universities.  Some of the 
results in this paper have been derived using the HEALPix 
\citep{gorskietal05} package.

\begin{table*}
\begin{center}
\caption{Highest SNR combination of wavelengths for a representative selection of spectral resolutions for the Earth-Moon system set at 1.34 parsecs from the observer using a 12-m coronagraphic space telescope exposed for 24 hours.\label{table03}
}
\begin{tabular}{lcccccccccc}
R      & SNR   & Signal & $\lambda_{1}$ & $\Delta\lambda_{1}$ & Moon Flux & Planet Flux & $\lambda_{2}$  & $\Delta\lambda_{2}$ & Moon Flux & Planet Flux    \\
($\lambda/\Delta\lambda$) & ($\sigma$)  & (mas)& ($\mu$m) & ($\mu$m)& \% of Total & \% of Total &($\mu$m) & ($\mu$m) & \% of Total & \% of Total\\ 
    \hline\\
    
1.0 &  17.86 & 0.1264 &  0.295  &  0.147-0.442 & 0.75 & 99.25 &  1.998  &  0.999-2.997 & 7.34 & 92.66 \\
1.5 &  19.30 & 0.2138 &  0.328  &  0.219-0.438 & 0.74 & 99.26 &  2.010  &  1.340-2.681 & 11.89 & 88.11 \\
2.0 &  18.56 & 0.4082 &  0.344  &  0.258-0.430 & 0.72 & 99.28 &  2.390  &  1.792-2.987 & 22.01 & 77.99 \\
3.0 &  15.55 & 0.1673 &  0.378  &  0.315-0.441 & 0.70 & 99.30 &  1.616  &  1.347-1.886 & 9.43 & 90.57 \\
4.0 &  16.20 & 0.8316 &  0.340  &  0.298-0.383 & 0.60 & 99.40 &  2.665  &  2.332-2.999 & 43.96 & 56.04 \\
5.0 &  17.89 & 1.3296 &  0.348  &  0.313-0.383 & 0.56 & 99.44 &  2.726  &  2.454-2.999 & 69.89 & 30.11 \\
10.0 &  14.62 & 0.6805 &  0.341  &  0.324-0.358 & 0.48 & 99.52 &  1.924  &  1.828-2.021 & 35.96 & 64.04 \\
15.0 &  13.00 & 0.7828 &  0.345  &  0.334-0.357 & 0.46 & 99.54 &  1.907  &  1.844-1.971 & 41.29 & 58.71 \\
20.0 &  12.09 & 0.9061 &  0.347  &  0.338-0.355 & 0.47 & 99.53 &  1.922  &  1.874-1.970 & 47.72 & 52.28 \\
30.0 &  10.61 & 1.0492 &  0.344  &  0.338-0.349 & 0.45 & 99.55 &  1.933  &  1.900-1.965 & 55.17 & 44.83 \\
40.0 &  9.50 & 1.1002 &  0.351  &  0.347-0.356 & 0.46 & 99.54 &  1.935  &  1.911-1.959 & 57.84 & 42.16 \\
50.0 &  8.47 & 1.1089 &  0.348  &  0.345-0.351 & 0.47 & 99.53 &  1.939  &  1.920-1.958 & 58.29 & 41.71 \\
60.0 &  7.86 & 1.1152 &  0.349  &  0.346-0.352 & 0.49 & 99.51 &  1.947  &  1.931-1.963 & 58.64 & 41.36 \\
100.0 &  6.35 & 1.1622 &  0.347  &  0.346-0.349 & 0.46 & 99.54 &  1.949  &  1.940-1.959 & 61.07 & 38.93 \\
150.0 &  5.28 & 1.1572 &  0.349  &  0.348-0.350 & 0.45 & 99.55 &  1.955  &  1.949-1.962 & 60.80 & 39.20 \\
200.0 &  4.73 & 1.1759 &  0.348  &  0.347-0.349 & 0.57 & 99.43 &  1.931  &  1.926-1.936 & 61.89 & 38.11 \\    
    \end{tabular}
\end{center}
\end{table*}

\begin{table*}
\begin{center}
\caption{Highest SNR combination of wavelengths for a representative selection of spectral resolutions for the Jovian-Earth system set at 10 parsecs from the observer using a 12-m coronagraphic space telescope exposed for 24 hours.\label{table04}
}
\begin{tabular}{lcccccccccc}
R      & SNR   & Signal & $\lambda_{1}$ & $\Delta\lambda_{1}$ & Moon Flux & Planet Flux & $\lambda_{2}$  & $\Delta\lambda_{2}$ & Moon Flux & Planet Flux    \\
($\lambda/\Delta\lambda$) & ($\sigma$)  & (mas)& ($\mu$m) & ($\mu$m)& \% of Total & \% of Total &($\mu$m) & ($\mu$m) & \% of Total & \% of Total\\ 
    \hline\\
1.0 &  3.33 & 0.0088 &  0.858  &  0.429-1.286 & 0.54 & 99.46 &  1.288  &  0.644-1.931 & 0.97 & 99.03 \\
2.0 &  11.99 & 0.0450 &  0.571  &  0.428-0.714 & 0.33 & 99.67 &  0.947  &  0.710-1.184 & 2.53 & 97.47 \\
3.0 &  15.07 & 0.0775 &  0.514  &  0.428-0.600 & 0.25 & 99.75 &  0.858  &  0.715-1.001 & 4.03 & 95.97 \\
4.0 &  16.59 & 0.1282 &  0.490  &  0.429-0.552 & 0.21 & 99.79 &  0.887  &  0.777-0.998 & 6.47 & 93.53 \\
5.0 &  16.53 & 0.1680 &  0.476  &  0.428-0.523 & 0.20 & 99.80 &  0.899  &  0.809-0.989 & 8.40 & 91.60 \\
10.0 &  16.01 & 0.3812 &  0.452  &  0.429-0.474 & 0.18 & 99.82 &  0.925  &  0.879-0.971 & 18.79 & 81.21 \\
20.0 &  12.64 & 0.3119 &  0.456  &  0.445-0.468 & 0.17 & 99.83 &  0.882  &  0.860-0.904 & 15.40 & 84.60 \\
30.0 &  13.49 & 0.5779 &  0.451  &  0.443-0.458 & 0.17 & 99.83 &  0.894  &  0.879-0.908 & 28.38 & 71.62 \\
40.0 &  15.55 & 1.0089 &  0.449  &  0.443-0.454 & 0.16 & 99.84 &  0.892  &  0.881-0.903 & 49.42 & 50.58 \\
50.0 &  15.64 & 1.2301 &  0.452  &  0.447-0.456 & 0.16 & 99.84 &  0.891  &  0.882-0.900 & 60.22 & 39.78 \\
60.0 &  14.73 & 1.2646 &  0.455  &  0.451-0.459 & 0.16 & 99.84 &  0.890  &  0.882-0.897 & 61.91 & 38.09 \\
80.0 &  13.25 & 1.3377 &  0.448  &  0.445-0.451 & 0.15 & 99.85 &  0.890  &  0.884-0.896 & 65.47 & 34.53 \\
100.0 &  12.20 & 1.3499 &  0.447  &  0.445-0.450 & 0.15 & 99.85 &  0.887  &  0.883-0.892 & 66.06 & 33.94 \\
120.0 &  11.04 & 1.3296 &  0.453  &  0.451-0.455 & 0.14 & 99.86 &  0.887  &  0.883-0.891 & 65.06 & 34.94 \\
150.0 &  9.86 & 1.3327 &  0.470  &  0.468-0.471 & 0.13 & 99.87 &  0.890  &  0.887-0.893 & 65.21 & 34.79 \\
200.0 &  8.53 & 1.3319 &  0.430  &  0.429-0.431 & 0.12 & 99.88 &  0.886  &  0.884-0.889 & 65.15 & 34.85 \\
250.0 &  7.68 & 1.3217 &  0.491  &  0.490-0.492 & 0.12 & 99.88 &  0.889  &  0.887-0.891 & 64.66 & 35.34 \\
300.0 &  7.22 & 1.3790 &  0.431  &  0.430-0.432 & 0.10 & 99.90 &  0.889  &  0.888-0.890 & 67.43 & 32.57 \\
400.0 &  6.32 & 1.4111 &  0.431  &  0.430-0.431 & 0.09 & 99.91 &  0.890  &  0.889-0.891 & 68.99 & 31.01
\end{tabular}
\end{center}
\end{table*}


\begin{thebibliography}{}

\bibitem[Agol(2007)]{agol2007} Agol, E., 2007, MNRAS, 374, 1271

\bibitem[Agnor 
\& Asphaug(2004)]{2004ApJ...613L.157A} Agnor, C., \& Asphaug, E.\ 2004, 
\apjl, 613, L157

\bibitem[Agnor 
\& Hamilton(2006)]{2006Natur.441..192A} Agnor, C.~B., \& Hamilton, D.~P.\ 
2006, \nat, 441, 192

\bibitem[Andrade-Ines \& Michtchenko (2014)]{AndradeInes2014} Andrade-Ines, E.\ \& Michtchenko, T.~A., 2014, MNRAS, 444, 2167

\bibitem[Armstrong et al.(2014)]{Armstrongetal2014} Armstrong, J., Barnes, R., Domagal-Goldman, S.,
Breiner, J., Quinn, T.R.\ \& Meadows, V.S., 2014, Astrobiology, 14, 277

\bibitem[{{Aumann} {et~al.}(2003){Aumann}, {Chahine}, {Gautier}, {Goldberg},
  {Kalnay}, {McMillin}, {Revercomb}, {Rosenkranz}, {Smith}, {Staelin}, {Strow},
  \& {Susskind}}]{aumann03}
{Aumann}, H.~H., {et~al.} 2003, IEEE Transactions on Geoscience and Remote
  Sensing, 41, 253
  
\bibitem[{Bailey} (1998)]{Bailey1998} Bailey, J.~A., 1998, SPIE, 3355, 932

\bibitem[{{Barnes} \& {O'Brien}(2002)}]{barneso'brien02}{Barnes}, J.~W., \& 
{O'Brien}, D.~P. 2002, \apj, 575, 1087

\bibitem[{{Beer} {et~al.}(2001){Beer}, {Glavich}, \& {Rider}}]{beeretal01}
{Beer}, R., {Glavich}, T.~A., \& {Rider}, D.~M. 2001, \ao, 40, 2356

\bibitem[{{Buratti} {et~al.}(2011){Buratti}, {Hicks}, {Nettles}, {Staid},
  {Pieters}, {Sunshine}, {Boardman}, \& {Stone}}]{burattietal11}
{Buratti}, B.~J., {Hicks}, M.~D., {Nettles}, J., {Staid}, M., {Pieters}, C.~M.,
  {Sunshine}, J., {Boardman}, J., \& {Stone}, T.~C. 2011, Journal of
  Geophysical Research (Planets), 116, E00G03

\bibitem[{{Burrows} \& {Sharp}(1999)}]{burrows&sharp99}
{Burrows}, A., \& {Sharp}, C.~M. 1999, \apj, 512, 843

\bibitem[{{Burrows} {et~al.}(2004){Burrows}, {Sudarsky}, \&
  {Hubeny}}]{burrowsetal04}
{Burrows}, A., {Sudarsky}, D., \& {Hubeny}, I. 2004, \apj, 609, 407

\bibitem[Cabrera 
\& Schneider(2007)]{2007A&A...464.1133C} Cabrera, J., \& Schneider, J.\ 
2007, \aap, 464, 1133 

\bibitem[Cameron 
\& Ward(1976)]{1976LPI.....7..120C} Cameron, A.~G.~W., \& Ward, W.~R.\ 1976, 
Lunar and Planetary Science Conference, 7, 120

\bibitem[Canup 
\& Ward(2002)]{2002AJ....124.3404C} Canup, R.~M., \& Ward, W.~R.\ 2002, 
\aj, 124, 3404

\bibitem[Cash(2006)]{cash2006} Cash, W., 2006, \nat, 442, 51

\bibitem[Cockell et al.(2009)]{Cockell2009} Cockell, C.S., et al., 2009,
Astrobiology, 9, 1

\bibitem[Schneider et al.(2015)]{Schneider2015} Schneider, J., Lainey, V.\ \& Cabrera, J., 2015, International Journal of Astrobiology, 14, 191

\bibitem[{{Chandrasekhar}(1960)}]{chandrasekhar60}
{Chandrasekhar}, S. 1960, {Radiative Transfer} (Dover)

\bibitem[{{Crisp}(1997)}]{crisp97}
{Crisp}, D. 1997, \grl, 24, 571

\bibitem[Dalcanton et al.(2015)]{Dalcanton2015} Dalcanton, J., Seager, S., Aigrain, S., Battel, S., Brandt, N., Conroy, C., Feinberg, L., Gezari, S., Guyon, O., Harris, W., Hirata, C., Mather, J., Postman, M., Redding, D., Schiminovich, D., Stahl, H.P.\ \& Tumlinson, J., 2015, arXiv:1507.04779, \url{http://www.hdstvision.org}

\bibitem[Day et al.(2003)]{Day2003} Day, P.K., LeDuc, H.G., Mazin, B.A., Vayonakis,
A., Zmuidzinas, J., 2003, Nature, 425, 817

\bibitem[Duncan \& Kipping (2013)]{2013MNRAS.432.2994F} Forgan, D.\ \& Kipping, D., 2013, Monthly Notices of the Royal Astronomical Society, 432, 2994

\bibitem[Eggl et al.(2013)]{Eggl2013} Eggl, S., Haghighipour, N.\ \& Pilat-Lohinger, E., 2013, ApJ, 764, 130

\bibitem[{{G{\'o}rski} {et~al.}(2005){G{\'o}rski}, {Hivon}, {Banday},
  {Wandelt}, {Hansen}, {Reinecke}, \& {Bartelmann}}]{gorskietal05}
{G{\'o}rski}, K.~M., {Hivon}, E., {Banday}, A.~J., {Wandelt}, B.~D., {Hansen},
  F.~K., {Reinecke}, M., \& {Bartelmann}, M. 2005, \apj, 622, 759

\bibitem[{Guyon} {et~al.}(2006)]{Guyon2006} Guyon, O., Pluzhnik, E.~A., Kuchner, M.~J., Collins, B., \& Ridgway, S.~T.,
2006, ApJS, 167, 81

    \bibitem[{Guyon} {et~al.}(2013)]{Guyon2013} {Guyon}, O. {et~al.},
2013, ApJ, 767, 11

\bibitem[Macintosh et al.(2006)]{Macintosh2006}  Macintosh, et al., 2006, SPIE,
6272, 62720

\bibitem[{{Hall} {et~al.}(1995){Hall}, {Riggs}, \& {Salomonson}}]{halletal95}
{Hall}, D.~K., {Riggs}, G., \& {Salomonson}, V.~V. 1995, Remote Sensing of
  Environment, 54, 127
  
\bibitem[Hartmann 
\& Davis(1975)]{1975Icar...24..504H} Hartmann, W.~K., \& Davis, D.~R.\ 
1975, {\it Icarus}, 24, 504

\bibitem[Heller(2012)]{2012A&A...545L...8H} Heller, R.\ 2012, \aap, 545, 
LL8

\bibitem[Heller 
\& Barnes(2013)]{2013AsBio..13...18H} Heller, R., \& Barnes, R.\ 2013, 
Astrobiology, 13, 18

\bibitem[Heller(2014)]{Heller2014} Heller, R., 2014, \apj, 787, 14

\bibitem[Heller et al.\ (2014)]{2014AsBio..14..798H} Heller, R., Williams,
D., Kipping, D., Limbach, M.~A., Turner, E., Greenberg, R., Sasaki, T.,
Bolmont, \'E, Grasset, O., Lewis, K., Barnes, R.\ \& Zulaga, J.~I., 2014, 
Astrobiology, 14, 798

\bibitem[Heller \& Armstrong(2014)]{HellerArmstrong2014} Heller, R.\ \& Armstrong,
J., 2014, Astrobiology, 14, 50

\bibitem[Heller \& Albrecht(2014)]{HellerAlbrecht2014} Heller, R.\ \& Albrecht,
S., 2014, \apjl, 796, L1

\bibitem[Heller 
\& Barnes(2015)]{2015IJAsB..14..335H} Heller, R., \& Barnes, R.\ 2015, 
International Journal of Astrobiology, 14, 335

\bibitem[Heller \& Pudritz (2015a)]{2015A&A...578A..19H} Heller, R.\ \& Pudritz, R., 2015, Astronomy \& Astrophysics, 578, A19

\bibitem[Heller \& Pudritz (2015b)]{2015ApJ...806..181H} Heller, R.\ \& Pudritz, R., 2015, \apj, 806, 181

\bibitem[Hinkel \& Kane (2013)]{2013ApJ...774...27H} Hinkel, N.~R.\ \& Kane, S.~R., 2013, The Astrophysical Journal, 774, 27

\bibitem[Kaltenegger(2000)]{2000ESASP.462..199K} Kaltenegger, L.\ 2000, 
Exploration and Utilisation of the Moon, 462, 199

\bibitem[Kaltenegger(2010)]{2010ApJ...712L.125K} Kaltenegger, L.\ 2010, 
The Astrophysical Journal, 712, 125

\bibitem[Kipping(2009)]{2009MNRAS.392..181K} Kipping, D.~M.\ 2009, \mnras, 
392, 181

\bibitem[Kipping(2011)]{2011MNRAS.416..689K} Kipping, D.~M.\ 2011, \mnras, 
416, 689

\bibitem[Kipping et al.(2012)]{2012ApJ...750..115K} Kipping, D.~M., 
Bakos, G.~{\'A}., Buchhave, L., Nesvorn{\'y}, D., \& Schmitt, A.\ 2012, 
\apj, 750, 115

\bibitem[Kouveliotou et al.(2014)]{2014arXiv1401.3741K} Kouveliotou, C., 
Agol, E., Batalha, N., et al.\ 2014, arXiv:1401.3741 

\bibitem[{{Lane} \& {Irvine}(1973)}]{lane&irvine73}
{Lane}, A.~P., \& {Irvine}, W.~M. 1973, \aj, 78, 267

\bibitem[Laskar et al.(1993)]{1993Natur.361..615L} Laskar, J., Joutel, F., 
\& Robutel, P.\ 1993, \nat, 361, 615

\bibitem[Lawson et al.(2008)]{Lawson2008} Lawson, J.R., et al., 2008,
SPIE, 7013, 70132N

\bibitem[Lissauer et al.(2012)]{Lissauer2012} Lissauer, J.J., Barnes, J.W.\ \&
Chambers, J.E., 2012, Icarus, 217, 77

\bibitem[{{Livengood} {et~al.}(2011){Livengood},  {Deming}, {A'Hearn},
  {Charbonneau}, {Hewagama}, {Lisse}, {McFadden}, {Meadows}, {Robinson}, {Seager},
  {Wellnitz}}]{Livengood2011}
{Livengood}, T.~A., {et~al.} 2011, Astrobiology, 11, 907

\bibitem[Lunine 
\& Stevenson(1982)]{1982Icar...52...14L} Lunine, J.~I., \& Stevenson, 
D.~J.\ 1982, {\it Icarus}, 52, 14

\bibitem[Mazin et al. (2012)]{Mazin2012} Mazin, B., Bumble, B., 
Meeker, S.R., O'Brien, K., McHugh, S., Langman, E., 2012, Optics Express, 20, 1503

\bibitem[McCord(1966)]{1966AJ.....71..585M} McCord, T.~B.\ 1966, \aj, 71, 
585

\bibitem[McKinnon(1984)]{1984Natur.311..355M} McKinnon, W.~B.\ 1984, \nat, 
311, 355

\bibitem[{{Meadows} \& {Crisp}(1996)}]{meadows&crisp96}
{Meadows}, V.~S., \& {Crisp}, D. 1996, \jgr, 101, 4595

\bibitem[Moskovitz et al.(2009)]{2009AsBio...9..269M} Moskovitz, N.~A., 
Gaidos, E., \& Williams, D.~M.\ 2009, Astrobiology, 9, 269

\bibitem[Mosqueira 
\& Estrada(2003)]{2003Icar..163..198M} Mosqueira, I., \& Estrada, 
P.~R.\ 2003, {\it Icarus}, 163, 198 

\bibitem[{{Muinonen} {et~al.}(1989){Muinonen}, {Lumme}, {Peltoniemi}, \&
  {Irvine}}]{muinonenetal89}
{Muinonen}, K., {Lumme}, K., {Peltoniemi}, J., \& {Irvine}, W.~M. 1989, \ao,
  28, 3051
  
\bibitem[Ogihara 
\& Ida(2012)]{2012ApJ...753...60O} Ogihara, M., \& Ida, S.\ 2012, \apj, 753, 60

\bibitem[Peacock et al.(1996)]{Peacock1996} Peacock, A., Verhoeve, P.,
Rando, N., van Dordrecht, A., Taylor, B.G., Erd, C., Perryman, M.A.C.,
Venn, R., Howlett, J., Goldie, D.J., Lumley, J.\ \& Wallis, M., 1996, Nature,
381, 135

\bibitem[Peters 
\& Turner(2013)]{2013ApJ...769...98P} Peters, M.~A., \& Turner, E.~L.\ 2013, 
\apj, 769, 98

\bibitem[{{Peters} {et~al.}(2007){Peters}, {Jacobson}, {Sweeney}, {Andrews},
  {Conway}, {Masarie}, {Miller}, {Bruhwiler}, {Petron}, {Hirsch}, {Worthy},
  {van der Werf}, {Randerson}, {Wennberg}, {Krol}, \& {Tans}}]{petersetal07}
{Peters}, W., {et~al.} 2007, Proceedings of the National Academy of Science,
  1041, 18925
  
\bibitem[Rein et al.(2014)]{2014PNAS..111.6871R} Rein, H., Fujii, Y., 
\& Spiegel, D.~S.\ 2014, Proceedings of the National Academy of Science, 111, 6871 

\bibitem[Reynolds et al.(1987)]{Reynolds1987} Reynolds, R.T., McKay, C.P.\
\& Kasting, J.F., 1987, Advances in Space Research, 7, 125

\bibitem[{{Riggs} {et~al.}(1999){Riggs}, {Hall}, \& {Ackerman}}]{riggsetal95}
{Riggs}, G., {Hall}, D.~K., \& {Ackerman}, S.~A. 1999, Remote Sensing of
  Environment, 68, 152

\bibitem[Robinson(2011)]{2011ApJ...741...51R} Robinson, T.~D.\ 2011, \apj, 
741, 51

\bibitem[{{Robinson} {et~al.}(2010){Robinson}, {Meadows}, \&
  {Crisp}}]{robinsonetal10}
{Robinson}, T.~D., {Meadows}, V.~S., \& {Crisp}, D. 2010, \apjl, 721, L67

\bibitem[{{Robinson} {et~al.}(2011){Robinson}, {Meadows}, {Crisp}, {Deming},
  {A'Hearn}, {Charbonneau}, {Livengood}, {Seager}, {Barry}, {Hearty},
  {Hewagama}, {Lxisse}, {McFadden}, \& {Wellnitz}}]{robinsonetal11}
{Robinson}, T.~D., {et~al.} 2011, Astrobiology, 11, 393

\bibitem[Robinson et al.(2014)]{robinsonetal14} Robinson, T.~D., 
Ennico, K., Meadows, V.~S., et al.\ 2014, \apj, 787, 171 

\bibitem[Sartoretti 
\& Schneider(1999)]{1999A&AS..134..553S} Sartoretti, P., \& Schneider, J.\ 
1999, \aaps, 134, 553

\bibitem[{{Salomonson} {et~al.}(1989){Salomonson}, {Barnes}, {Maymon},
  {Montgomery}, \& {Ostrow}}]{salomonsonetal89}
{Salomonson}, V.~V., {Barnes}, W.~L., {Maymon}, P.~W., {Montgomery}, H.~E., \&
  {Ostrow}, H. 1989, IEEE Transactions on Geoscience and Remote Sensing, 27,
  145
  
\bibitem[Sasaki et al.(2010)]{Sasaki2010} Sasaki, T., Stewart, G.R.\ \& Ida, S.,
2010, \apj, 714, 1052
  
\bibitem[Scharf(2006)]{2006ApJ...648.1196S} Scharf, C.~A.\ 2006, \apj, 648, 
1196

\bibitem[Simon et 
al.(2007)]{2007A&A...470..727S} Simon, A., Szatm{\'a}ry, K., \& Szab{\'o}, 
G.~M.\ 2007, \aap, 470, 727

\bibitem[Spergel et al.(2015)]{2015arXiv50303757S} Spergel, D., et al.,
2015, arXiv:1503.03757

\bibitem[Stapelfeldt et al.(2014)]{2014SPIE.9143E..2KS} Stapelfeldt,
K.R., et al., 2014, Proc.\ SPIE 9143, 91432K

\bibitem[Stark et al.(2015)]{Stark2015} Stark, C.~C., et al., 2015,
arXiv:1506.01723

\bibitem[{{Sudarsky} {et~al.}(2003){Sudarsky}, {Burrows}, \&
  {Hubeny}}]{sudarskyetal03}
{Sudarsky}, D., {Burrows}, A., \& {Hubeny}, I. 2003, \apj, 588, 1121

\bibitem[Szab{\'o} et 
al.(2006)]{2006A&A...450..395S} Szab{\'o}, G.~M., Szatm{\'a}ry, K., 
Div{\'e}ki, Z., \& Simon, A.\ 2006, \aap, 450, 395

\bibitem[Tusnski 
\& Valio(2011)]{2011ApJ...743...97T} Tusnski, L.~R.~M., \& Valio, A.\ 
2011, \apj, 743, 97

\bibitem[{{Waters} {et~al.}(2006){Waters}, {Froidevaux}, {Harwood}, {Jarnot},
  {Pickett}, {Read}, {Siegel}, {Cofield}, {Filipiak}, {Flower}, {Holden},
  {Lau}, {Livesey}, {Manney}, {Pumphrey}, {Santee}, {Wu}, {Cuddy}, {Lay},
  {Loo}, {Perun}, {Schwartz}, {Stek}, {Thurstans}, {Boyles}, {Chandra},
  {Chavez}, {Chen}, {Chudasama}, {Dodge}, {Fuller}, {Girard}, {Jiang}, {Jiang},
  {Knosp}, {Labelle}, {Lam}, {Lee}, {Miller}, {Oswald}, {Patel}, {Pukala},
  {Quintero}, {Scaff}, {Vansnyder}, {Tope}, {Wagner}, \&
  {Walch}}]{watersetal06}
{Waters}, J.~W., {et~al.} 2006, IEEE Transactions on Geoscience and Remote
  Sensing, 44, 1075

\bibitem[Ward et al.(2002)]{2002LPI....33.2017W} Ward, W.~R., Agnor, C.~B., 
\& Canup, R.~M.\ 2002, Lunar and Planetary Science Conference, 33, 2017

\bibitem[Whelan 
\& Garcia(2008)]{2008LNP...742..123W} Whelan, E., \& Garcia, P.\ 2008, 
Jets from Young Stars II, 742, 123
  
\bibitem[Williams 
\& Knacke(2004)]{2004AsBio...4..400W} Williams, D.~M., \& Knacke, R.~F.\ 
2004, Astrobiology, 4, 400

\bibitem[Williams et al.(1997)]{1997Natur.385..234W} Williams, D.~M., 
Kasting, J.~F., \& Wade, R.~A.\ 1997, \nat, 385, 234

\end{thebibliography}
\end{document}